\documentclass[]{pasj02} 
\usepackage[switch,mathlines]{lineno} 

\jyear{2024}
\Received{}
\Accepted{}

\begin{document} 

\title{Study of the origin of the azimuthal variation of synchrotron X-ray spectrum from SNR RX~J0852.0$-$4622}

\author{
 Dai \textsc{Tateishi},\altaffilmark{1}\altemailmark\orcid{0000-0003-0248-4064} \email{d.tateishi.astro@gmail.com} 
 Nobuaki \textsc{Sasaki},\altaffilmark{2} \orcid{0000-0001-8138-7697}
 Yukikatsu \textsc{Terada},\altaffilmark{2,3} \orcid{0000-0002-2359-1857}
 Satoru \textsc{Katsuda},\altaffilmark{2} \orcid{0000-0002-1104-7205}
 Shiu-Hang \textsc{Lee},\altaffilmark{4} \orcid{0000-0002-2899-4241}
 Hidetoshi \textsc{Sano},\altaffilmark{5} \orcid{0000-0003-2062-5692}
 Yasuo \textsc{Fukui},\altaffilmark{6} \orcid{0000-0002-8966-9856}
 and 
 Aya \textsc{Bamba}\altaffilmark{1,7,8}\orcid{0000-0003-0890-4920}
}
\altaffiltext{1}{Department of Physics, Graduate School of Science, The University of Tokyo, 7-3-1 Hongo, Bunkyo-ku, Tokyo 113-0033, Japan}
\altaffiltext{2}{Graduate School of Science and Engineering, Saitama University, 255 Shimo-Okubo, Sakura-ku, Saitama, 338-8570, Japan}
\altaffiltext{3}{Japan Aerospace Exploration Agency, Institute of Space and Astronautical Science, Sagamihara, Kanagawa, Japan}
\altaffiltext{4}{Department of Astronomy, Kyoto University Oiwake-cho, Kitashirakawa, Sakyo-ku, Kyoto 606-8502, Japan}
\altaffiltext{5}{Faculty of Engineering, Gifu University, 1-1 Yanagido, Gifu, Gifu 501-1193, Japan}
\altaffiltext{6}{Department of Physics, Nagoya University, Furo-cho, Chikusa-ku, Nagoya, Aichi 464-8601, Japan}
\altaffiltext{7}{Research Center for the Early Universe, School of Science, The University of Tokyo, 7-3-1 Hongo, Bunkyo-ku, Tokyo 113-0033, Japan}
\altaffiltext{8}{Trans-Scale Quantum Science Institute, The University of Tokyo, 7-3-1 Hongo, Bunkyo-ku, Tokyo 113-0033, Japan}



\maketitle

\begin{abstract}
We report the azimuthal distribution of the X-ray energy spectrum of non-thermal dominant supernova remnant RX~J0852.0$-$4622.
The X-rays from the shock region observed by the X-ray astronomy satellite Suzaku/XIS in the energy range of 2–8 keV are well described by the absorbed power-law model and can be parameterized with flux and photon index.
The X-ray flux and photon index are bimodally distributed in relation to the azimuthal angle.
To understand its origin, we examined three possible causes: azimuthal variation by (1) the galactic magnetic field, (2) cloud density, and (3) shock velocity.
From the polarization observations of stars near the SNR, we find that the Galactic magnetic field around the SNR is not aligned.
This result leads us to conclude that the azimuthal variation of the X-ray spectrum is most likely not caused by the Galactic magnetic field.
The X-ray fluxes are positively correlated with the cloud density with a significance of $\sim 5\sigma$, and the azimuthal distributions of these physical quantities are particularly pronounced in the northern part of the SNR.
In addition, the X-ray fluxes on the southern part of the SNR are positively correlated with the shock velocity.
This phenomenon can be qualitatively explained by the increase in roll-off energy due to the amplification of the magnetic field by (A) the interaction between the shock and dense clouds in the north and (B) the fast shock velocity in the south of the SNR.
Since the shock velocity is likely related to the cloud density interacting with the shock, we conclude that the azimuthal variation of cloud density most likely causes the azimuthal variations of the X-ray flux and photon index.
\end{abstract}


\section{Introduction}
\label{sec:intro}
RX~J0852.0-4622 (also known as Vela Jr. and G266.2-1.2) is a shell-type supernova remnant (SNR) discovered by the ROSAT All-Sky Survey \citep{Aschenbach_1998}.
The angular diameter is about $1.9^{\circ}$ \citep{Camilloni_2023}, and its age and distance to the SNR are estimated as $\sim$2,000--5,000 years old and $\sim$1 kpc, respectively, based on X-ray and gamma-ray observations \citep{Aschenbach_1998, Iyudin_1998, Aschenbach_1999, Tsunemi_2000, Slane_2001, Kargaltsev_2002, Katsuda_2008, Katsuda_2009, Acero_2013, Allen_2015, Camilloni_2023}.

This SNR is observed in wideband wavelengths from radio \citep{Combi_1999, Duncan_2000, Reynoso_2006} to very-high-energy (VHE) gamma-rays \citep{Acero_2016, Katagiri_2005, Enomoto_2006, Aharonian_2005, Aharonian_2007, HESS_2018_velajr}.
In the X-ray band, synchrotron radiation from relativistic electrons dominates, whereas thermal radiation is feeble \citep{Aschenbach_1998, Aschenbach_1999, Tsunemi_2000, Slane_2001, Aharonian_2007, Bamba_2005, Pannuti_2010, Allen_2015, Hiraga_2009, Fukui_2023, Iyudin_2005, Katsuda_2008, Katsuda_2009, Acero_2013, Kishishita_2013, Takeda_2016, Camilloni_2023}.
This result indicates that the SNR is located in a region with low ISM density \citep{Slane_2001}.
The energy spectrum of non-thermal emission from X-ray to gamma-ray can be well explained using both leptonic and hadronic models under reasonable physical conditions of acceleration \citep{HESS_2018_velajr}.
By comparing the X-ray flux, gamma-ray flux, and cloud density, \citet{Fukui_2024} evaluated the contribution of leptonic and hadronic components to the observed VHE gamma-rays to $52\%\,\pm\,1\%:48\%\,\pm\,1\%$.

A neutron star, AX~J0851.9$-$4617 (also known as CXOU~J085201.4$-$461753 and 2XMM~J085201.4$-$461753), is reported close to the geometric center of the SNR (e.g., \cite{Slane_2001}).
The surface temperature, radius, and luminosity obtained by X-ray spectrum analysis are similar to those of the Central Compact Objects (CCOs), a subclass of young neutron stars.
This indicates that the progenitor of this SNR is a massive star, causing a core-collapsed supernova.

The non-thermal X-ray emission of this SNR shows a limb-brightened morphology, with the northern
and southern edges brighter, and is reminiscent of remnants of bipolar non-thermal morphology such as SN 1006.
However, the origin of this distribution is still in debate.
By comparing the distribution of VHE gamma-ray and clouds, \citet{Fukui_2017} and \citet{Fukui_2024} concluded that variation in cloud density causes the azimuthal variation.
On the other hand, the angle between the shock and the Galactic magnetic field may also create these azimuthal variations, as studied in the shell-type SNR SN 1006 \citep{Rothenflug_2004, Petruk_2011, Zhou_2023}.
The roll-off energy of synchrotron radiation from SNRs are proportional to the square of the velocity of shock in the synchrotron loss-limited cases \citep{Zirakashvili_2007}.
Therefore, shock velocity may also contribute to the azimuthal variation of synchrotron X-ray flux, as in the northern part of SNR RCW 86 (e.g., \cite{Yamaguchi_2008}) and several young SNRs (e.g., \cite{Bamba_2005b}).

The purpose of this study is to investigate the origin of the azimuthal variation in X-rays.
The section proceeds with section \ref{sec:velajr_observation} describing the observations and data reduction, section \ref{sec:velajr_analysis} analysis and results, section \ref{sec:discussion} providing some discussions, and ends with a summary.
Throughout this paper, we used \citet{Lodders_2009} for the solar abundance ratio.

\section{Observation and data reduction}
\label{sec:velajr_observation}
We analyzed Suzaku/XIS archival data observed between 2005 and 2013.
We used data products from the Suzaku pipeline processing version 3.0.22.44 with calibration versions XIS: 20151005 and XRT: 20110630, software package HEASoft\footnote{http://heasarc.gsfc.nasa.gov/ftools} 6.34 \citep{heasoft_2014} for the analysis, and XSPEC v12.14.1 for the spectral analysis.
Suzaku/XIS consists of 4 CCDs: 3 front-illuminated (FI) CCDs (XIS0, 2, and 3) and a back-illuminated (BI) CCD (XIS1).
XIS data were screened using standard criteria, i.e., we discarded events with Earth elevation below $5^{\circ}$, the sun-irradiated-Earth elevation below $20^{\circ}$, or when the spacecraft was in an orbit phase within $436$ s after the South Atlantic Anomaly.
In the analysis, we accepted only the XIS events with standard grades (0, 2, 3, 4, and 6).
The region IDs, coordinates of the region center, azimuthal angle, observation IDs, observation date, and effective exposure are listed in table \ref{tab:velajr_suzaku}.
\begin{figure*}
    \begin{center}
        \includegraphics[width=16cm,clip]{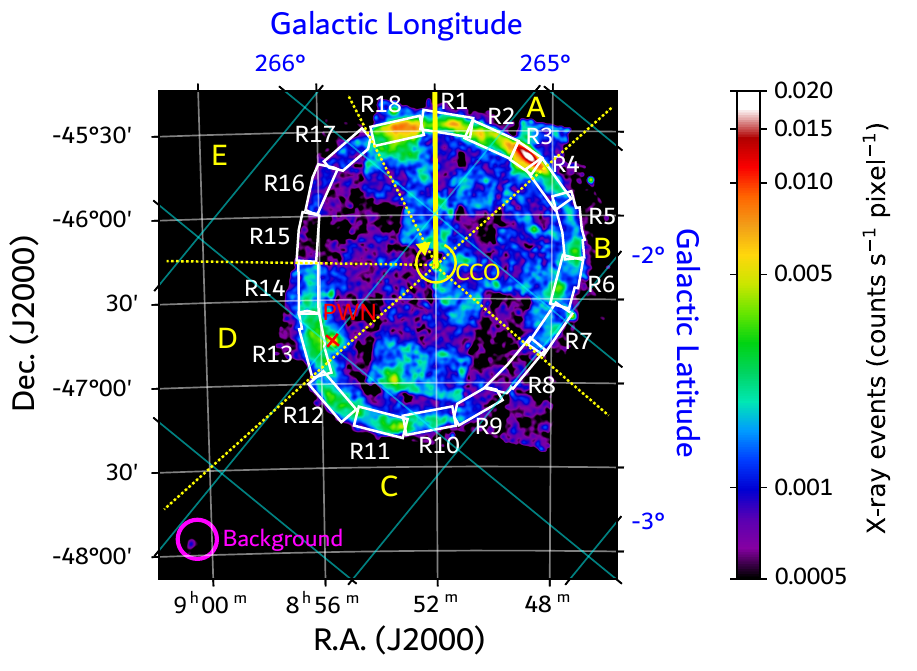}
    \end{center}
    \caption{
    Spectral extracted regions for source and background data are overlaid on the X-ray (2.0--5.0 keV) image of SNR RX~J0852.0$-$4622 observed by Suzaku/XIS.
    The source and background regions are indicated by white boxes and magenta circle, respectively.
    The white and cyan grid lines represent the equatorial coordinate system (J2000.0) and Galactic coordinate, respectively.
    The yellow vertical line denotes the origin of the azimuthal angle, which increases clockwise.
    {Alt text: X-ray image of SNR RX~J0852.0$-$4622 and spectrum extraction region of our analysis.} 
    }
    \label{fig:velajr_extract_reg}
\end{figure*}

\begin{longtable}{ccccccc}
  \caption{List of analyzed Suzaku/XIS archival data}\label{tab:velajr_suzaku}  
\hline\noalign{\vskip3pt} 
  Region ID & R.A.  & Dec. & Azimuthal angle \footnotemark[$*$] & Obs. ID & Obs. date & Exposure  \\   [2pt] 
            & (J2000.0) & (J2000.0) & (deg.) &  & (YYYY-MM-DD) & (ks)  \\  [2pt] 
\hline\noalign{\vskip3pt} 
\endfirsthead      
\hline\noalign{\vskip3pt} 
  Region ID & R.A.  & Dec. & Azimuthal angle \footnotemark[$*$] & Obs. ID & Obs. date & Exposure  \\  [2pt] 
\hline\noalign{\vskip3pt} 
\endhead
\hline\noalign{\vskip3pt} 
\endfoot
\hline\noalign{\vskip3pt} 
\multicolumn{2}{@{}l@{}}{\hbox to0pt{\parbox{160mm}{\footnotesize
\hangindent6pt\noindent
\hbox to6pt{\footnotemark[$*$]\hss}\unskip%
  We defined $0^{\circ}$ of the azimuthal angle as the yellow vertical line of figure \ref{fig:velajr_extract_reg}, and increased in the clockwise direction. 
}\hss}} 
\endlastfoot 
R1         &  8:51:35.7140  & -45:27:23.516 & 354.5--15    & 502027010 & 2007-07-05  &  10.98   \\ \hline
R2         &  8:50:04.5450  & -45:33:08.209 & 13.5--36     & 502026010 & 2007-07-05  &  10.97   \\
           &                &               &              & 500010010 & 2005-12-19  &  175.5   \\\hline
R3         &  8:48:54.9705  & -45:39:51.615 &  36--47      & 500010010 & 2005-12-19  &  175.5   \\
           &                &               &              & 502024010 & 2007-07-04  &  10.87   \\
           &                &               &              & 502025010 & 2007-07-04  &  10.30   \\\hline
R4         &  8:48:07.7630  & -45:48:57.749 & 45--61       & 502023010 & 2007-07-04  &  10.66   \\
           &                &               &              & 502024010 & 2007-07-04  &  10.87   \\\hline
R5         &  8:47:14.6715  & -46:03:57.287 & 61--87.5     & 502035010 & 2007-07-09  &  9.409   \\
           &                &               &              & 508037010 & 2013-11-23  &  29.20   \\
           &                &               &              & 508038010 & 2013-11-24  &  34.96   \\\hline
R6         &  8:47:22.6171  & -46:24:35.158 & 87.5--109    & 502036010 & 2007-07-09  &  11.04   \\
           &                &               &              & 508038010 & 2013-11-24  &  34.96   \\\hline
R7         &  8:48:03.990   & -46:40:46.024 & 108.5--130.5 & 503036010 & 2008-07-05  &  12.37   \\\hline
R8         &  8:49:18.6499  & -46:57:13.464 & 129--158.5   & 503037010 & 2008-07-06  &  13.16   \\
           &                &               &              & 503046010 & 2008-07-09  &  11.04   \\\hline
R9         &  8:50:37.4271  & -47:07:36.100 & 154--173.5   & 503049010 & 2008-07-09  &  12.55   \\\hline
R10        &  8:52:15.3900  & -47:13:32.997 & 173--191.5   & 503048010 & 2008-07-09  &  11.14   \\\hline
R11        &  8:53:57.6500  & -47:13:57.722 & 191--209     & 503045010 & 2008-07-08  &  12.02   \\
           &                &               &              & 508062010 & 2013-12-03  &  27.36   \\\hline
R12        &  8:55:38.4350  & -47:04:32.600 & 209.5--228   & 508061010 & 2013-11-24  &  14.47   \\
           &                &               &              & 508062010 & 2013-12-03  &  27.36   \\\hline
R13        &  8:56:16.1610  & -46:43:10.199 & 225--251     & 503042010 & 2008-07-07  &  10.12   \\
           &                &               &              & 503043010 & 2008-07-08  &  11.23   \\
           &                &               &              & 508060010 & 2013-11-25  &  40.61   \\\hline
R14        &  8:56:24.0830  & -46:25:42.792 & 248.5--271.5 & 503041010 & 2008-07-07  &  11.38   \\\hline
R15        &  8:56:25.6499  & -46:06:57.574 & 272--293     & 503031010 & 2008-07-03  &  19.82   \\\hline
R16        &  8:55:56.9957  & -45:49:52.320 & 292.5--314.5 & 502030010 & 2007-07-06  &  13.22   \\\hline
R17        &  8:54:54.8940  & -45:37:22.963 & 312.5--333   & 502029010 & 2007-07-05  &  14.58   \\\hline
R18        &  8:53:16.9430  & -45:29:52.407 & 334--354.5   & 502028010 & 2007-07-05  &  11.53   \\\hline
Background &  9:00:31.2210  & -47:54:12.663 & N/A          & 500010020 & 2005-12-23  &  59.22   \\
\end{longtable}

\section{Spectrum analysis and results}
\label{sec:velajr_analysis}

\subsection{Spectrum extraction}
Figure \ref{fig:velajr_extract_reg} shows the 2.0-5.0 keV X-ray image of the SNR obtained by Suzaku/XIS.
The brightness of 2.0-5.0 keV X-rays is enhanced in the north and south of the SNR, showing a bimodal distribution.
To determine the spectral extraction region, we first estimated the shock front from the X-ray image (figure \ref{fig:velajr_extract_reg}) by eye.
We then extracted the spectrum from regions up to $7'$ downstream of the shock front, as represented by the white boxes in figure \ref{fig:velajr_extract_reg}.
We extracted events from XIS 0, 1, 2 (Obs. ID 500010010 only) and 3, and merged the spectrum obtained from the FI CCDs.
To quantitatively evaluate the azimuthal variation, we defined $0^{\circ}$ of the azimuthal angle as the yellow vertical line of figure \ref{fig:velajr_extract_reg}, and increased in the clockwise direction.
The azimuth angles for each region are summarized in table \ref{tab:velajr_suzaku}.

\subsection{Background estimation}
\label{sec:velajr_background}
To estimate the contribution of the X-ray background, we analyzed the Suzaku/XIS data (Obs. ID: 500010020. The details of this observation are presented in table \ref{tab:velajr_suzaku}).
We selected this observation because there were no bright X-ray-emitting objects inside the field-of-view, and the mean Galactic latitude was approximately the same as the average of the SNR, making it an appropriate region to estimate the Galactic X-ray background emission.
We extracted the spectrum from a circular region with a radius of $7'$, as shown in the magenta circle in figure \ref{fig:velajr_extract_reg}, and subtracted the non-X-ray background (NXB) created by \texttt{xisnxbgen} \citep{Tawa_2008}.
We considered the solar wind charge exchange (SWCX) and/or local hot bubble (LHB), cosmic X-ray background (CXB), and Galactic ridge X-ray emission (GRXE) as X-ray astrophysical background components and modeled them using the following procedure (e.g., \cite{Kaneda_1997, Uchiyama_2009b, Kataoka_2013, Mizuno_2015}).
\begin{itemize}
    \item Contribution of SWCX and/or LHB\\
        We approximated the X-ray spectrum of SWCX and/or the LHB with the collisionally-ionized thermal X-ray model ($\mathrm{APEC}$) with a temperature and abundance fixed at 0.1 keV and solar abundance, respectively (e.g., \cite{Fujimoto_2007}).
    \item Contribution of CXB\\
        We modeled it using the canonical spectral model of CXB, i.e., the absorbed (TBabs; \cite{tbabs}) power-law model ($\mathrm{TBabs}\, \times \, \mathrm{powerlaw}$) with a photon index fixed to 1.4 \citep{Moretti_2003}.
    \item Contribution of GRXE\\
        The GRXE is known to vary from place to place on the Galactic plane and strongly depends on the Galactic latitude.
        We modeled it with the canonical spectral model of GRXE, i.e., the absorbed two-temperature collisionally-ionized thermal X-ray model (e.g., \cite{Kaneda_1997, Uchiyama_2009b, Mizuno_2015}) ($\mathrm{TBabs}\, \times\, \mathrm{APEC(low)} + \mathrm{TBabs}\, \times\, \mathrm{APEC(high)}$).
\end{itemize}

\begin{figure}
    \begin{center}
        \includegraphics[width=8cm,clip]{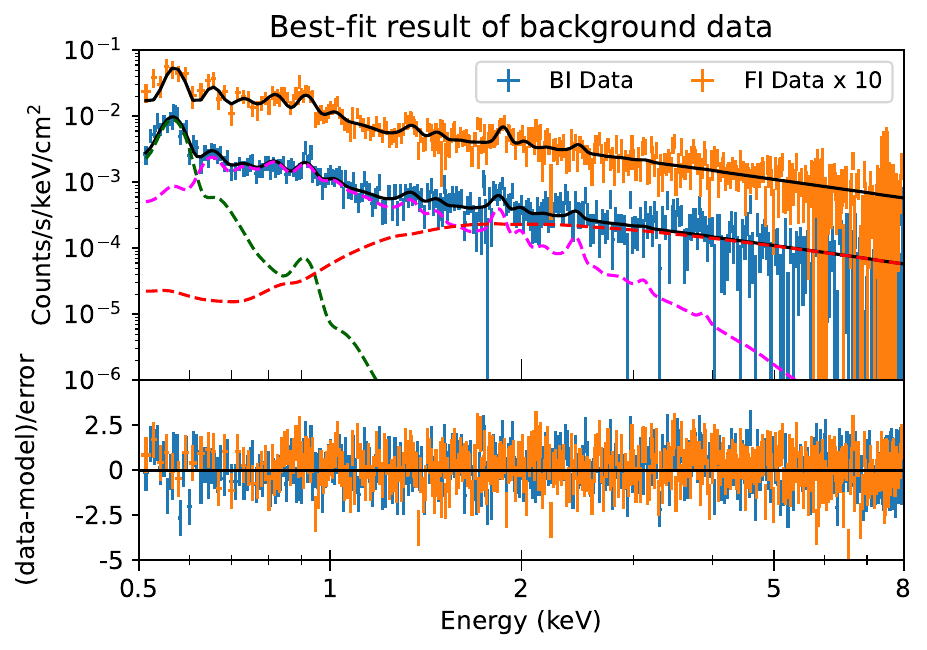}
    \end{center}
    \caption{
    Best-fit results of the modeling of the X-ray astrophysical background spectrum.
    The blue and orange data points represent the NXB-subtracted X-ray spectrum observed by the BI CCD and FI CCDs, respectively.
    The green, red, and magenta dashed lines represent the contribution of SWCX and/or LHB, CXB, and GRXE for the X-ray spectrum of BI CCD, respectively.
    The black lines represent the total best-fit model.
    {Alt text: A line graph. 
    The x-axis shows the energy from 0.5 to 8 kilo-electron volts. The y-axis shows the count from 0.000001 to 0.1 counts per second per kilo-electron volt per square centimeter in the upper part and the residuals of minus 5 to 5 in the lower part.}
    }
    \label{fig:velajr_x-ray_back}
\end{figure}

The overall expression of the X-ray astrophysical background model is $\mathrm{APEC(SWCX\,and/or\, LHB}) + \mathrm{TBabs(GRXE, low)}\, \times\, \mathrm{APEC(GRXE, low)} + \mathrm{TBabs(GRXE, high)}\, \times\, \mathrm{APEC(GRXE, high)} + \mathrm{TBabs(CXB)}\, \times\, \mathrm{powerlaw}$.
We then fitted the NXB-subtracted background X-ray spectrum using this model to estimate their contribution.
The best-fit for the modeling of the astrophysical background is shown in figure \ref{fig:velajr_x-ray_back}, and the associated best-fit parameters are given in table \ref{tab:velajr_x-ray_back_para}. 
This background model is then used for all the following XIS spectral analyses.
We note that an additional X-ray astrophysical background component may exist due to the presence of the Vela SNR in the line-of-sight direction.
See section \ref{sec:vela_contamination} for details.

\begin{figure*}
    \begin{center}
        \includegraphics[width=16cm,clip]{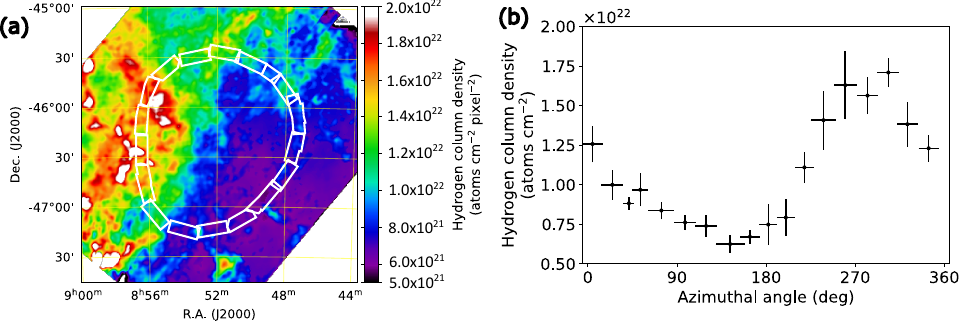}
    \end{center}
    \caption{
    The image and azimuthal distribution of foreground hydrogen column density derived from radio $^{12}$CO and H{\sc i} observations (see appendix \ref{sec:AppendixB} for details of the analysis).
    {Alt text: Two graphs.
    Figure (a) is an image of the foreground hydrogen column density of the SNR and spectrum extraction regions of our analysis.
    The x-axis and y-axis of the figure (b) shows the azimuthal angle of minus 6 to 360 degrees, and foreground hydrogen column density from 5 times 10 to the power of 21 to 2 times 10 to the power of 22, respectively}.
    }
    \label{fig:hydrogen_cloumn_density}
\end{figure*}

\begin{table*}
\tbl{The best-fit parameters of the X-ray astrophysical background spectrum obtained by Suzaku/XIS (0.5-8 keV)\footnotemark[$*$]}{
\begin{tabular}{clc}
\hline
Model & Parameter & Value \\
\hline
APEC (SWCX and/or LHB) & Plasma temperature (keV)                                                       & 0.1 (Fixed)           \\
                       & Abundance $(Z/Z_{\odot})$                                                      & 1 (Fixed)             \\
                       & 2-8 keV Flux $(\times\, 10^{-23}\, {\rm erg\,cm^{-2}\,s^{-1}\,arcmin^{-2}})$   & $5.0\pm0.5$           \\ \hline
TBabs (GRXE, Low)      & Absorbing column density $(\times 10^{21}\, {\rm atoms\,cm^{-2}})$              & $4.4_{-3.4}^{+1.7}$   \\ \hline
APEC (GRXE, Low)       & Plasma temperature (keV)                                                       & $0.25_{-0.02}^{+0.03}$\\
                       & Abundance $(Z/Z_{\odot})$                                                      & 0.4 (Fixed)           \\
                       & 2-8 keV Flux $(\times\, 10^{-18}\,{\rm erg\,cm^{-2}\,s^{-1}\,arcmin^{-2}})$    & $3.7\pm0.3$           \\ \hline
TBabs (GRXE, High)     & Absorbing column density $(\times\, 10^{22}\,{\rm atoms\,cm^{-2}})$             & $1.2\pm0.4$           \\ \hline
APEC (GRXE, High)      & Plasma temperature (keV)                                                       & $0.9\pm0.1$           \\
                       & Abundance $(Z/Z_{\odot})$                                                      & $< 0.4$               \\
                       & 2-8 keV Flux $(\times\, 10^{-16}\,{\rm erg\,cm^{-2}\,s^{-1}\,arcmin^{-2}})$     & $3.6\pm0.3$           \\ \hline
TBabs (CXB)            & Absorbing column density $(\times\, 10^{22}\,{\rm atoms\,cm^{-2}})$             & $1.5_{-0.7}^{+1.2}$   \\ \hline
Power-law (CXB)        & Photon index                                                                   & 1.4 (Fixed)           \\
                       & 2-8 keV Flux $(\times\, 10^{-15}\,{\rm erg\,cm^{-2}\,s^{-1}\,arcmin^{-2}})$    & $4.1\pm0.2$           \\ \hline
                       & $\chi^{2}$/d.o.f.                                                              & 0.91 (=654.39/722)    \\ \hline
\end{tabular}}
\label{tab:velajr_x-ray_back_para}  
\begin{tabnote}
\footnotemark[*] The errors represent a 90\% confidence level on an interesting single parameter. 
\end{tabnote}
\end{table*}

\subsection{Spectrum fitting and results}
\subsubsection{Spectrum fitting}
\label{sec:Spectrum_fitting}
In the following, we only focused on X-rays within an energy range of 2--8 keV to avoid contamination by the thermal X-rays below $\sim 1.4$ keV of Vela SNR in the line-of-sight direction \citep{Camilloni_2023}.
See section \ref{sec:vela_contamination} for details.

In order to evaluate the shape and flux of the synchrotron X-ray spectrum quantitatively, we fitted the NXB-subtracted spectrum with an absorbed power-law model (${\rm TBabs\, \times\, powerlaw}$), in addition to the contribution of the astrophysical background model.
The photon index cannot be accurately determined via X-ray spectral analysis above 2 keV due to its degeneracy with absorbing column density.
To disentangle the degeneracy, we fixed the absorbing column density to the foreground proton column density in the line-of-sight direction obtained by radio observation.
The azimuthal distribution of proton column density to the SNR is shown in figure \ref{fig:hydrogen_cloumn_density} (See appendix \ref{sec:AppendixB} for details of the analysis).
We simultaneously fitted the spectrum obtained from the FI and BI CCDs with the model described above.

The resultant best-fit NXB-subtracted spectrum is shown for region R1 in figure \ref{fig:velajr_spec_bestfit}, and is provided in appendix \ref{sec:X-ray_spectrum_list} for regions R2 to R16 (figure \ref{fig:appendix_velajr_2to8keV_1}) and regions R17 and R18 (figure \ref{fig:appendix_velajr_2to8keV_2}).
We found that the NXB-subtracted X-ray spectrum exceeded more than a few times that of the X-ray background in all regions.
The best-fit parameters are shown in table \ref{tab:velajr_bestfit_para}.
The reduced $\chi^2$ value in each region ($\sim1$) indicates that an absorbed power-law model can explain the excess X-ray spectrum well.

\begin{figure}
    \begin{center}
        \includegraphics[width=8cm,clip]{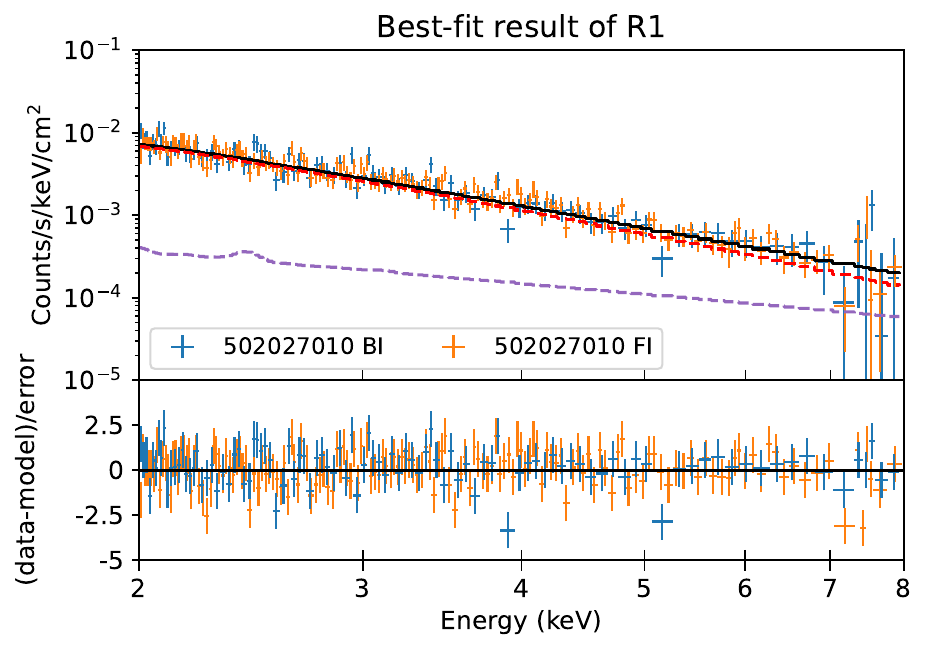}
    \end{center}
    \caption{
    The best-fit results of the NXB-subtracted X-ray spectrum.
    The blue and orange data points represent the NXB-subtracted 2-8 keV X-ray spectrum in region R1 observed by the BI CCD and FI CCDs, respectively.
    The purple and red dashed lines represent the X-ray background (GRXE + CXB + SWCX) and absorbed power-law components, respectively, for the BI CCD.
    The black lines represent the total best-fit model.
    {Alt text: A line graph. 
    The x-axis shows the energy from 2 to 8 kilo-electron volts. 
    The y-axis shows the count from 0.00001 to 0.1 counts per second per kilo-electron volt per square centimeter in the upper part and the residuals of minus 5 to 5 in the lower part.}
    }
    \label{fig:velajr_spec_bestfit}
\end{figure}

\begin{table*}
\tbl{Best-fit parameters of the 2-8 keV spectra along the rim of SNR RX J0852.0-4622\footnotemark[$*$]}{
\begin{tabular}{ccccc}
\hline
  Region ID & Foreground hydrogen column density\footnotemark[$**$] & Photon index & 2--8 keV X-ray flux & $\chi^{2}$/d.o.f. \\ 
  & (${\rm atoms\, cm^{-2}}$, fixed) && ($\rm erg\, cm^{-2}\, s^{-1}\, arcmin^{-2}$) & \\    
\hline
R1  & $(1.3 \pm 0.2) \times 10^{22}$   & $3.20\pm0.09$          & $(3.4 \pm 0.1)\times 10^{-14}$   & 0.97 (=255.1/262) \\
R2  & $(1.0 \pm 0.1) \times 10^{22}$   & $3.00\pm0.02$          & $(5.42 \pm 0.04)\times 10^{-14}$ & 1.12 (=2556.14/2272) \\
R3  & $(8.8 \pm 0.4) \times 10^{21}$   & $2.84\pm0.01$          & $(8.24 \pm 0.04)\times 10^{-14}$ & 1.04 (=2988.75/2729) \\
R4  & $(9.7 \pm 1.1) \times 10^{21}$   & $2.98\pm0.09$          & $(2.39 \pm 0.07)\times 10^{-14}$ & 1.21 (=429.51/354) \\
R5  & $(8.4 \pm 0.6) \times 10^{21}$   & $3.13\pm0.07$          & $(1.42 \pm 0.04)\times 10^{-14}$ & 1.30 (=842.89/647) \\
R6  & $(7.6 \pm 0.5) \times 10^{21}$   & $3.11\pm0.08$          & $(1.67 \pm 0.05)\times 10^{-14}$ & 1.03 (=472.47/460) \\
R7  & $(7.4 \pm 0.7) \times 10^{21}$   & $3.3\pm0.2$            & $(1.21 \pm 0.07)\times 10^{-14}$ & 1.11 (=177.61/160) \\
R8  & $(6.3 \pm 0.6) \times 10^{21}$   & $3.1\pm0.2$            & $(8.5 \pm 0.7)\times 10^{-15}$   & 0.99 (=120.36/121) \\
R9  & $(6.7 \pm 0.5) \times 10^{21}$   & $3.0\pm0.2$            & $(1.16 \pm 0.07)\times 10^{-14}$ & 0.92 (=122.85/134) \\
R10 & $(7.5 \pm 1.3) \times 10^{21}$   & $2.8\pm0.1$            & $(2.50 \pm 0.09)\times 10^{-14}$ & 0.96 (=215.27/224) \\
R11 & $(7.9 \pm 1.2) \times 10^{21}$   & $2.76_{-0.07}^{+0.08}$ & $(2.73 \pm 0.08)\times 10^{-14}$ & 1.38 (=564.29/409) \\
R12 & $(1.1 \pm 0.1) \times 10^{22}$   & $2.93\pm0.08$          & $(2.19 \pm 0.06)\times 10^{-14}$ & 0.96 (=359.64/373) \\
R13 & $(1.5 \pm 0.1) \times 10^{22}$   & $2.77\pm0.05$          & $(2.91 \pm 0.05)\times 10^{-14}$ & 0.96 (=799.60/830) \\
R14 & $(1.6 \pm 0.3) \times 10^{22}$   & $3.0\pm0.2$            & $(1.19 \pm 0.8)\times 10^{-14}$  & 0.88 (=130.55/149) \\
R15 & $(1.6 \pm 0.2) \times 10^{22}$   & $2.7\pm0.3$            & $(5.2 \pm 0.6)\times 10^{-15}$   & 1.36 (=190.38/140) \\
R16 & $(1.71 \pm 0.09) \times 10^{22}$ & $3.2_{-0.4}^{+0.5}$    & $(4.5 \pm 0.7)\times 10^{-15}$   & 1.21 (=90.69/75) \\
R17 & $(1.4 \pm 0.2) \times 10^{22}$   & $3.1\pm0.2$            & $(9.6 \pm 0.6)\times 10^{-15}$   & 0.96 (=151.18/158) \\
R18 & $(1.23 \pm 0.09) \times 10^{22}$ & $2.95\pm0.08$          & $(3.6 \pm 0.1)\times 10^{-14}$   & 0.94 (=285.17/304) \\
\hline
\end{tabular}}\label{tab:velajr_bestfit_para}  
\begin{tabnote}
\footnotemark[*] The errors represent a 90\% confidence level on an interesting single parameter.  \\ 
\footnotemark[**] Foreground hydrogen column densities were estimated from radio observation.  The errors represent a $1\sigma$ confidence level.\\
\end{tabnote}
\end{table*}

\subsubsection{Contamination by thermal X-rays of Vela SNR}
\label{sec:vela_contamination}
Vela SNR is located in the line-of-sight direction of the SNR RX~J0852.0$-$4622.
The X-ray emission from this SNR is dominated by thermal emission below 1.4 keV \citep{Camilloni_2023}, and the presence of X-ray synchrotron radiation has not been confirmed.
However, in regions with weak synchrotron X-rays from SNR RX~J0852.0$-$4622, the thermal emission of the Vela SNR may affect the photon index and flux of the non-thermal emission in 2--8 keV energy band.
To quantitatively evaluate this effect, we fit the X-ray spectrum in the energy band of 0.5--8 keV of four regions with weak synchrotron X-ray emission (R6, 8, 12, and 16) with an absorbed thermal model and an absorbed power-law model ($\rm TBabs \times vpshock + TBabs \times powerlaw$).
We used the background models described in section \ref{sec:velajr_background}.

The best-fit parameters are given in table \ref{tab:with_velaSNR}, and the best-fit spectrum of region R6 is shown in figure \ref{fig:with_velaSNR_R7}, and in figures \ref{fig:appendix_velajr_05to8keV} of appendix \ref{sec:appendix_05to80_fit_result} for regions R8, R12, and R16.
The reduced $\chi^2$ value in each region ($\sim 1$) indicates that the model can explain the excess X-ray spectrum well.
The temperature and ionization timescale of the thermal component obtained by the analysis were roughly consistent with the values observed in SGR/eROSITA \citep{Camilloni_2023}.
The X-ray fluxes and photon indexes obtained for these four regions agree within error with the results of table \ref{tab:velajr_bestfit_para} from the 2-8 keV analysis (see section \ref{sec:Spectrum_fitting}).
This result leads us to conclude that the effect of thermal X-rays from the Vela SNR is negligible to the X-ray spectrum in the energy range of 2--8 keV.

\begin{figure}
    \begin{center}
        \includegraphics[width=8.5cm,clip]{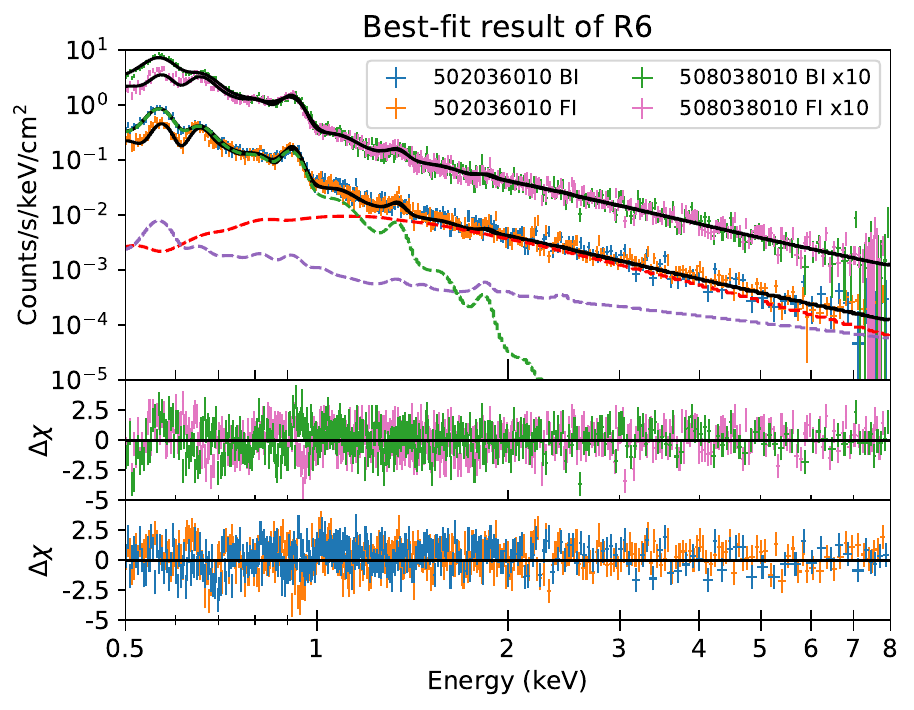}
    \end{center}
    \caption{
        The best-fit result of the NXB-subtracted X-ray spectrum (0.5-8.0 keV) obtained from region R6.
        The purple, red, and green dashed lines represent the X-ray background, absorbed power-law of SNR RX~J0852.0$-$4622, and absorbed thermal X-ray of Vela SNR, respectively, for X-ray spectrum obtained by BI CCD of Obs. ID of 502036010 (blue data points).
        {Alt text: A line graph. 
        The x-axis shows the energy from 0.5 to 8 kilo-electron volts. 
        The y-axis shows the count from 0.00001 to 10 counts per second per kilo-electron volt per square centimeter in the upper part and the residuals of minus 5 to 5 in the lower part.}
    }
    \label{fig:with_velaSNR_R7}
\end{figure}

\subsubsection{Contamination by nearby pulsar wind nebula PSR J0855-4644}
In the vicinity of region R13, there is a pulsar wind nebula (PWN) PSR J0855$-$4644 (see figure \ref{fig:velajr_extract_reg}).
This PWN is $<900$ pc from Earth and is not related to the SNR RX J0862.0-4622 \citep{Acero_2013}.
XMM-Newton/EPIC observations have shown that the X-ray synchrotron radiation from this PWN extends with a radius of $\sim 150''$ \citep{Acero_2013}.
To quantitatively evaluate the contribution of the PWN to the X-ray flux, we analyzed the X-ray spectrum and image obtained by Chandra/ACIS (Obs. ID: 13780 and 18640) and simulated its contribution using simulation tool \texttt{xissim}.
Our simulation revealed that the contamination of the PWN on the X-ray flux was $\sim 2\%$ in region R13, the closest spectrum extraction region from the PWN.
The result leads us to conclude that the contamination of the PWN is negligible to our results.

\begin{table*}
\tbl{The best-fit parameters of the X-ray spectrum of regions R6, 8, 12, and 16 in the energy range of 0.5--8 keV\footnotemark[$*$]}{
\begin{tabular}{clcccc}
\hline
  Model & Parameter & R6 & R8 & R12 & R16 \\     
\hline
TBabs\footnotemark[$**$]  & H column density $({\rm atoms\,cm^{-2}})$          & $(3.1\pm0.2)\times10^{21}$         & $(2.7\pm0.5) \times 10^{21}$           & $< 8.2 \times 10^{21}$                & $2.0_{-0.4}^{+0.7}\times10^{21}$  \\ \hline
VPSHOCK             & Plasma temperature (keV)                                 & $0.153_{-0.003}^{+0.005}$          & $0.17\pm0.01$                          & $0.28_{-0.04}^{+0.07}$                & $0.15_{-0.02}^{+0.01}$            \\
                    & $\rm (N(=O)/H)/(N(=O)/H)_{\odot}$                        & $0.31_{-0.03}^{+0.05}$             & $0.4\pm0.1$                            & $0.3_{-0.1}^{+0.3}$                   & $0.3_{-0.1}^{+0.2}$               \\
                    & $\rm (Ne/H)/(Ne/H)_{\odot}$                              & $0.40_{-0.04}^{+0.06}$             & $0.4\pm0.1$                            & $0.9_{-0.4}^{+0.7}$                   & $0.5\pm0.2$                       \\
                    & $\rm (Mg/H)/(Mg/H)_{\odot}$                              & $0.49_{-0.07}^{+0.09}$             & $0.3_{-0.1}^{+0.2}$                    & $<6.2$                                & $1.1_{-0.4}^{+0.8}$               \\
                    & $\rm (Fe/H)/(Fe/H)_{\odot}$                              & $1.7_{-0.2}^{+1.0}$                & $1.4_{-0.8}^{+1.5}$                    & $<3.7$                                & $0.6_{-0.5}^{+0.8}$               \\
                    & Ionization timescale $\tau\, \rm (s\,cm^{-3})$           & $3.2_{-0.4}^{+9.5}\times 10^{12}$  & $>1.9\times 10^{12}$                   & $1.4_{-0.6}^{+0.9}\times 10^{9}$     & $>3.8\times 10^{12}$              \\
                    & Red shift $\rm (km\,s^{-1})$                             & 0 (Fixed)                          & 0 (Fixed)                              & 0 (Fixed)                             & 0 (Fixed)                         \\
                    & 2-8 keV Flux $({\rm erg\,cm^{-2}\,s^{-1}\,arcmin^{-2}})$ & $2.0_{-0.1}^{+0.3}\times 10^{-17}$ & $1.3_{-0.3}^{+0.4}\times 10^{-17}$     & $1.7_{-0.9}^{+3.1}\times 10^{-17}$    & $(3\pm1)\times 10^{-18}$          \\ \hline
TBabs\footnotemark[$***$] & H column density $({\rm atoms\,cm^{-2}}, \rm Fixed)$     & $7.6\times 10^{21}$          & $6.3\times 10^{21}$                    & $1.1\times 10^{21}$                   & $1.71\times 10^{22}$              \\ \hline
Power-law           & Photon index                                             & $3.16\pm0.05$                      & $3.2\pm0.1$                            & $2.87\pm0.05$                         & $3.7_{-0.2}^{+0.3}$               \\
                    & 2-8 keV Flux $({\rm erg\,cm^{-2}\,s^{-1}\,arcmin^{-2}})$ & $(1.64\pm0.05)\times 10^{-14}$     & $8.4_{-0.6}^{+0.7}\times 10^{-15}$     & $(2.19\pm0.06)\times 10^{-14}$        & $4.2_{-0.6}^{+0.7}\times 10^{-15}$\\ \hline
                    & $\chi^{2}$/d.o.f.                                        & 2152.69/1706                       & 986.97/764                             & 884.69/913                            & 341.86/295                        \\ \hline
\end{tabular}}\label{tab:with_velaSNR}  
\begin{tabnote}
\footnotemark[*] The errors are the 90\% confidence level on an interesting single parameter.  \\ 
\footnotemark[**] Absorption for VPSHOCK component estimated from radio observation. \\
\footnotemark[***] Absorption for power-law component.\\
\end{tabnote}
\end{table*}

\subsection{Azimuthal distribution of parameters}
\label{sec:azimuth_dist_parameters}
\begin{figure*}
    \begin{center}
        \includegraphics[width=15cm,clip]{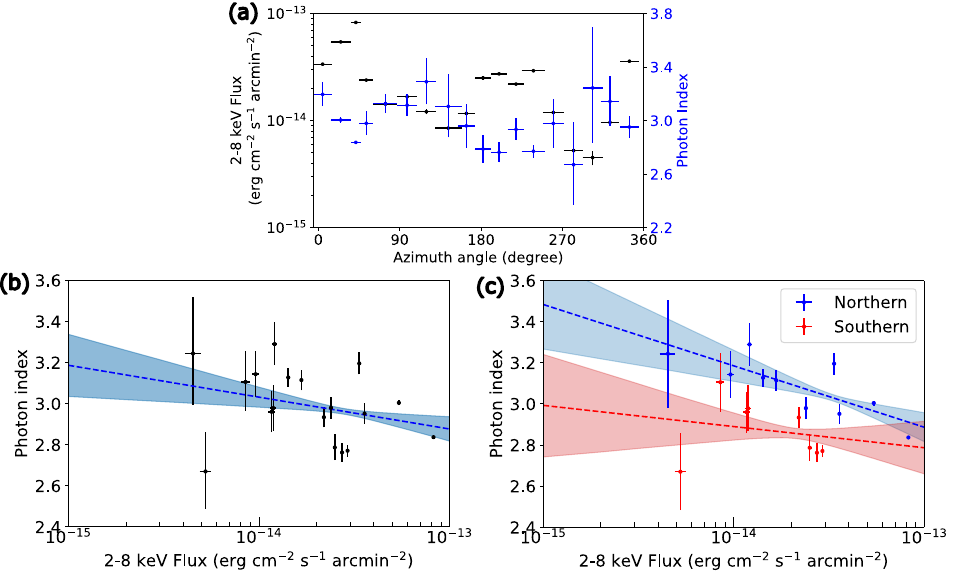}
    \end{center}
    \caption{
    Relation between the photon index and the X-ray flux.
    Figure (a) represents the azimuthal distribution of X-ray flux in the energy range of 2--8 keV (black) and photon index (blue) obtained from the rim of SNR RX~J0852.0$-$4622.
    Figure (b) represents the correlation plot of photon index and X-ray flux of the whole SNR.
    The best-fit results with a linear model ($\Gamma = a (\log F_{\rm X}-c) + b$) are represented in blue dashed lines, and the filled area represents 1$\sigma$ error.
    Figure (c) shows the correlation plot of the northern and southern regions, represented by blue and red dashed lines, respectively.
    We note that errors in figure (b) and (c) are $1\sigma$ confidence level.
    {Alt text: Three graphs.
    In figure (a), the x-axis shows the azimuthal angle of minus 6 to 360 degrees.
    The y-axis on the left shows the 2 to 8 kilo-electron volt X-ray flux from 10 to the power of minus 15 to 10 to the power of minus 13.
    The y-axis on the right shows the photon index from 2.2 to 3.8.
    In figures (b) and (c), the x-axis shows the 2 to 8 kilo-electron volt X-ray flux from 10 to the power of minus 15 to 10 to the power of minus 13. The y-axis shows the photon index from 2.4 to 3.6.
    }
    }
    \label{fig:flux_vs_azimuth}
\end{figure*}

We sorted the X-ray flux and photon index according to the azimuthal angle of each region.
The resultant azimuthal distributions are shown in figures \ref{fig:flux_vs_azimuth}.
From figure \ref{fig:flux_vs_azimuth} (a), the X-ray flux distributed bilaterally to the azimuthal angle and the peak intervals were $\sim180^{\circ}$ (maximized at $\sim 45^{\circ}$ and $\sim225^{\circ}$).
To a much lesser extent, the photon index profile suggests some bilateral variations, which appear anti-correlated with the flux profile.

Next, we assessed the correlation between (1) X-ray flux and photon index, (2) cloud density and X-ray spectrum, (3) gamma-ray counts and X-ray spectrum, (4) shock velocity and X-ray spectrum by fitting the various azimuthal profiles using linear or logarithmic functions.
In addition, the cause of the azimuthal distribution of the X-ray spectrum may be different between the northern and southern regions of the SNR.
To assess this scenario, we divided the SNR into northern and southern regions, as shown in table \ref{tab:def_north_south}, and evaluated the differences in correlations.

In the following analysis, we did not account for the intrinsic error of the observed data. 
Since the statistical error of X-ray flux is small compared to other observational data, intrinsic error may influence the correlation between parameters. 
For this reason, we performed an analysis that considered intrinsic error.
See appendix \ref{sec:fit_with_intscatter} for details.

\begin{table}
    \centering
    \caption{Definition of northern and southern regions}
    \begin{tabular}{ccc}
    \hline
                    & Angle (deg.)  & Region ID\\\hline
        Northern    & 292.5-130.5   & R1-R7, R16-R18 \\
        Southern    & 129--293      & R8--R15 \\\hline
    \end{tabular}
    \label{tab:def_north_south}
\end{table}

\subsubsection{Correlation between photon index and X-ray flux}
\label{sec:Xflux_index}
Figure  \ref{fig:flux_vs_azimuth} (b) shows the change in the photon index in response to variations in the X-ray flux.
We took the following steps to evaluate the correlation between the parameters quantitatively.
\begin{enumerate}
    \item We randomly assigned values for X-ray flux and photon index according to a Gaussian distribution and created 10,000 data sets.
    \item We generated 10,000 data sets with 18 data points of X-ray flux and photon index.
    The X-ray flux and photon index of each data follow a Gaussian distribution based on the median and error of the best-fit results of table \ref{tab:velajr_bestfit_para}.
    \item We fitted each data set with a logarithmic model ($\Gamma = a (\log F_{\rm X}-c) + b$, where $\Gamma$ and $F_{\rm X}$ represents the photon index and X-ray flux, respectively) and estimated the parameters $a$ and $b$. We used the parameter $c$ as the average of $F_{\rm X}$ of the data set and considered it during the fitting.
    \item We calculated the average of the best-fit results for parameter $a$ and its $1\sigma$ standard deviation $\Delta a$. Throughout the paper, we define the significance of correlation as $a/\Delta a$.
\end{enumerate}

The resultant significance of the anti-correlation is $1.47\sigma$, $2.10\sigma$, and $0.55\sigma$ for total, northern and southern region, respectively.
The best-fit parameters ($a, b$, and $c$) are presented in table \ref{tab:velajr_corelation}, and the best-fit results are shown in blue dashed lines of figure \ref{fig:flux_vs_azimuth} (b) for the total region, and blue and red dashed lines for northern and southern regions, respectively, of figure \ref{fig:flux_vs_azimuth} (c).

\subsubsection{Correlation between X-ray spectrum and cloud density}
\label{sec:X-ray_spectrum_and_cloud_density}
\begin{figure*}
\begin{center}
\includegraphics[width=15cm,clip]{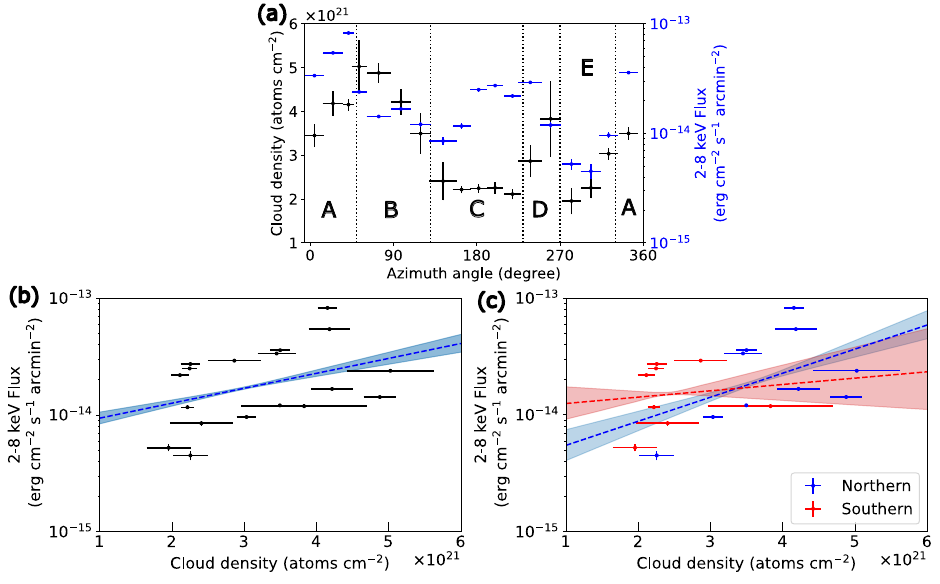}
\end{center}
\caption{
        Relation between the X-ray flux and the cloud density.
        Figure (a) represents the azimuthal distributions of the cloud density and the X-ray flux.
        Figure (b) represents the correlation plot with the best-fit results of a logarithm model ($\log F_{\rm X} = a (N_{\rm cloud}-c) + b$) (blue dashed line) and 1$\sigma$ error (blue filled area).
        Figure (c) shows the correlation plot of the northern and southern regions, represented by blue and red lines, respectively.
        We note that errors in figure (b) and (c) are $1\sigma$ confidence level.
        {Alt text: Three graphs.
        In figure (a), the x-axis shows the azimuthal angle of minus 6 to 360 degrees.
        The y-axis on the left shows the cloud density from 1 times 10 to the power of 21 to 6 times 10 to the power of 21 atoms per square centimeter.
        The y-axis on the right shows the 2 to 8 kilo-electron volt X-ray flux from 10 to the power of minus 15 to 10 to the power of minus 13.
        In figure (b) and (c) panels, the x-axis shows the cloud density from 1 times 10 to the power of 21 to 6 times 10 to the power of 21 atoms per square centimeter.
        The y-axis shows the 2 to 8 kilo-electron volt X-ray flux from 10 to the power of minus 15 to 10 to the power of minus 13.}
}
\label{fig:flux_vs_nh}
\end{figure*}

We assume the clouds with the line-of-sight velocity of 20 - 40 $\rm km\,s^{-1}$ estimated by radio observation \citep{Fukui_2024} as the clouds interacting with the shock of the SNR.
We used the proton column density map in figure 1 (c) of \citet{Fukui_2024} as the cloud density interacting with the SNR.
Figure \ref{fig:flux_vs_nh} (a), (b), and (c) shows the azimuthal distribution along the SNR rim of the X-ray flux and cloud density, the change in the X-ray flux in response to the cloud density for the total region, and northern and southern regions, respectively.
Figure \ref{fig:flux_vs_nh} (b) and (c) suggest that the X-ray flux and cloud density are correlated.
We fitted the data with a logarithmic model ($\log F_{\rm X} = a (N_{\rm cloud}-c) + b$, where $N_{\rm cloud}$ represents cloud density) and plotted the best-fit result.
The significance of the correlation is $5.12\sigma$, $4.10\sigma$, and $0.39\sigma$ for total, northern and southern regions, respectively.
The best-fit parameters ($a, b, c$) are presented in table \ref{tab:velajr_corelation}.

\begin{figure*}
\begin{center}
\includegraphics[width=15cm,clip]{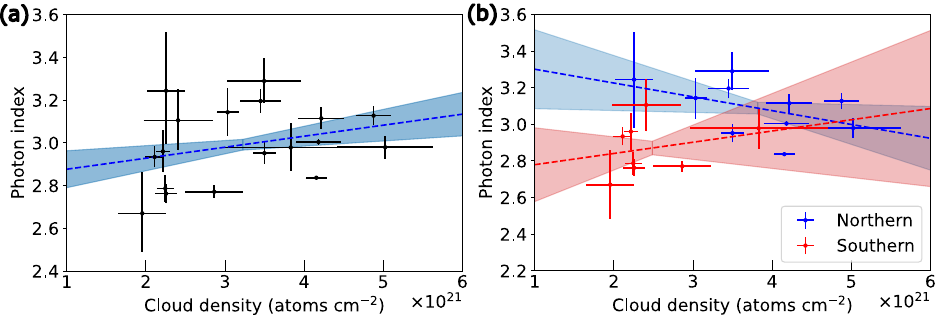}
\end{center}
\caption{
        Correlation plot of the photon index and the cloud density.
        The blue dashed lines and blue filled area in figure (a) represent the best-fit results of a linear model ($\Gamma = a (N_{\rm cloud}-c) + b$) and 1$\sigma$ error, respectively.
        Figure (b) shows the correlation plot of the northern and southern regions, represented by blue and red lines, respectively.
        We note that errors are $1\sigma$ confidence level.
        {Alt text: Two graphs.
        The x-axis shows the cloud density from 1 times 10 to the power of 21 to 6 times 10 to the power of 21 atoms per square centimeter.
        The y-axis shows the photon index from 2.4 to 3.6 and 2.2 to 3.6 in figure (a) and figure (b), respectively.}
}
\label{fig:index_vs_nh}
\end{figure*}

Figure \ref{fig:index_vs_nh} (a) and (b) shows a plot of the changes in the photon index in response to the cloud density for total, and northern and southern regions, respectively.
We fitted the plot with a linear model ($\Gamma = a (N_{\rm cloud}-c) + b$).
The best-fit results are shown in figure \ref{fig:index_vs_nh} and table \ref{tab:velajr_corelation}.
The significance of the correlation is $1.84\sigma$, $1.17\sigma$, and $0.54\sigma$ for total, northern and southern regions, respectively.

\subsubsection{Correlation between X-ray spectrum and gamma-ray counts}
\label{sec:x-ray_vs_gamma-ray}
\begin{figure}
    \begin{center}
        \includegraphics[width=8cm,clip]{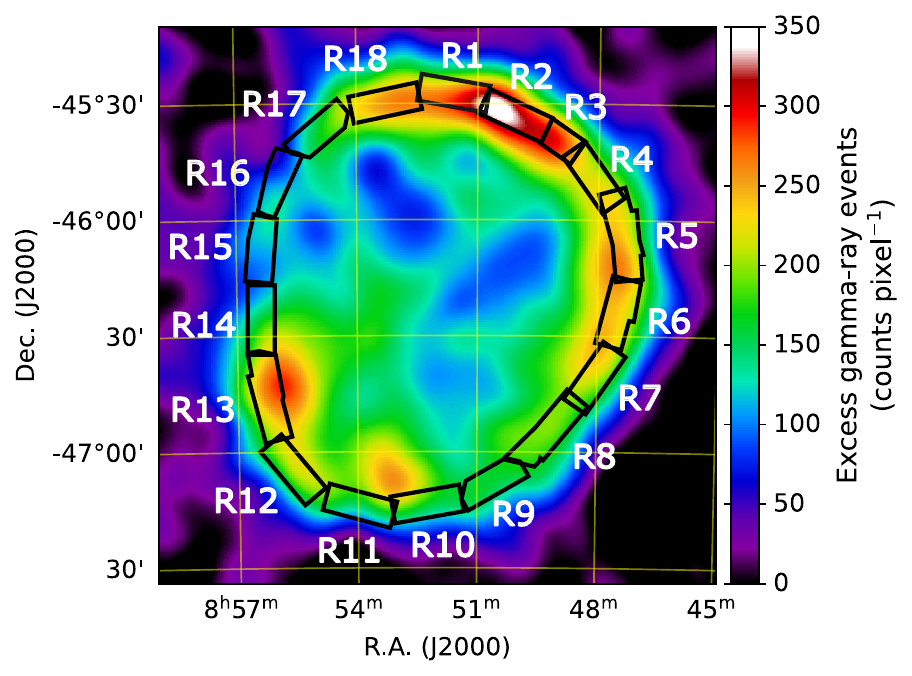}
    \end{center}
    \caption{
        An exposure-corrected gamma-ray image in the energy range of 100 GeV to 100 TeV obtained by H.E.S.S. \citep{HESS_2018_velajr}.
        The black boxes in the figure represent the spectrum extraction region.
        {Alt text: A gamma-ray image in the energy band of 100 Giga electron volt to 100 Tera electron volt of SNR RX~J0852.0$-$4622 and spectrum extraction region of our analysis.}
    }
    \label{fig:hess_gamma}
\end{figure}

\begin{figure*}
\begin{center}
\includegraphics[width=15cm,clip]{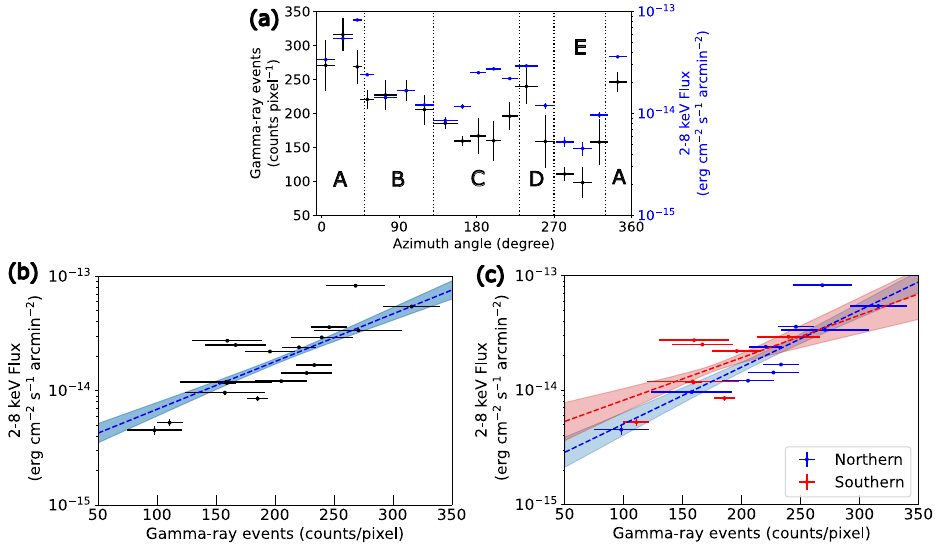}
\end{center}
\caption{
        Relation between X-ray flux and VHE gamma-ray counts (100 GeV to 100 TeV).
        Figure (a) represents the azimuthal distribution of the average excess gamma-ray counts (100 GeV - 100 TeV) observed by H.E.S.S..
        Figures (b) and (c) represent the correlation plot of average excess VHE gamma-ray counts (100 GeV to 100 TeV) and X-ray flux.
        The blue dashed lines and blue filled area in figure (b) represent the best-fit results of a linear model ($\log F_{\rm X} = a (F_{\rm \gamma}-c) + b$) and 1$\sigma$ error, respectively.
        Figure (c) shows the correlation plot of the northern and southern regions, represented by blue and red lines, respectively.
        We note that errors in figure (b) and (c) are $1\sigma$ confidence level.
        {Alt text: Three graphs.
        In figure (a), the x-axis shows the azimuthal angle of minus 6 to 360 degrees.
        The y-axis shows the average excess gamma-ray counts from 50 to 350 counts.
        In figures (b) and (c),} the x-axis shows the average excess gamma-ray counts from 50 to 350 counts per pixel.
        The y-axis shows the 2 to 8 kilo-electron volt X-ray flux from 10 to the power of minus 15 to 10 to the power of minus 13.
}
\label{fig:X-ray_vs_gamma-ray}
\end{figure*}

In order to evaluate the correlation level between X-ray and gamma-ray azimuthal profiles quantitatively, we first estimated the number of gamma-rays (100 GeV to 100 TeV) in each spectrum extraction region from the H.E.S.S. image (\cite{HESS_2018_velajr}).
The gamma-ray image and spectrum extraction regions are shown in figure \ref{fig:hess_gamma}.
The azimuthal profile of the VHE gamma-rays is shown in figure \ref{fig:X-ray_vs_gamma-ray} (a), and plot of the change in the X-ray flux in response to the gamma-ray excess counts is shown in figure \ref{fig:X-ray_vs_gamma-ray} (b) and (c) for total and northern and southern regions, respectively.
We fitted the data with a logarithmic model ($F_{\rm \gamma} = a (\log F_{\rm X}-c) + b$, where $F_{\rm \gamma}$ represents the gamma-ray excess counts) and plot the best-fit result.
The significance of the correlation is $7.92\sigma$, $6.62\sigma$ and $2.86\sigma$ for total, northern and southern regions, respectively.
The best-fit results and residuals are plotted in figure \ref{fig:X-ray_vs_gamma-ray} (b) and (c) for total and northern and southern regions, respectively, and best-fit parameters ($a, b, c$) are given in table \ref{tab:velajr_corelation}.
Note that we did not consider systematic errors related to the leakage of gamma-ray events between neighboring spectral extraction regions due to the size of the spectral extraction region, which is equivalent to that of the point spread function of the H.E.S.S. telescope.

\begin{figure*}
\begin{center}
\includegraphics[width=15cm,clip]{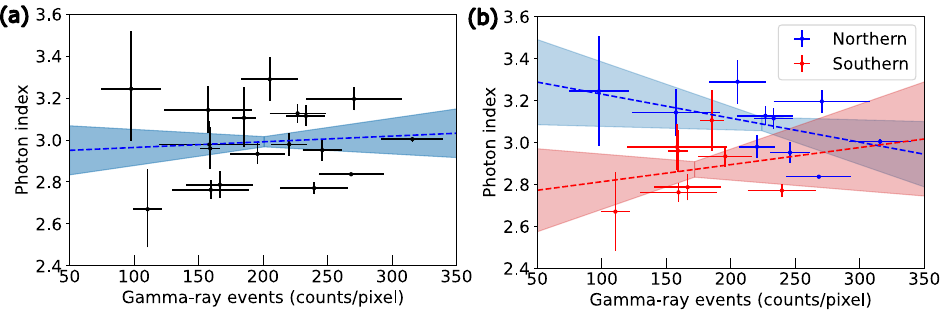}
\end{center}
\caption{
        Correlation plot of the photon index and the average excess VHE gamma-rays counts (100 GeV to 100 TeV).
        The blue dashed lines and blue filled area in figure (a) represent the best-fit results of a linear model ($\Gamma = a (F_{\rm \gamma}-c) + b$) and 1$\sigma$ error, respectively.
        Figure (b) shows the correlation plot of the northern and southern regions, represented by blue and red lines, respectively.
        We note that errors in both figures are $1\sigma$ confidence level.
        {Alt text: Two graphs.
        The x-axis shows the average excess gamma-ray counts from 50 to 350 counts per pixel.
        The y-axis shows the photon index from 2.4 to 3.6.}
}
\label{fig:index_vs_gamma-ray}
\end{figure*}

Figure \ref{fig:index_vs_gamma-ray} (a) and (b) shows a plot of the changes in the photon index in response to the gamma-ray flux, for total and northern and southern regions, respectively.
We fitted the plot with a linear model ($\Gamma = a (F_{\rm \gamma}-c) + b$).
The best-fit results are shown in figure \ref{fig:index_vs_gamma-ray} and table \ref{tab:velajr_corelation}.
The significance of the correlation is $0.44\sigma$, $1.18\sigma$, and $0.62\sigma$ for total, northern, and southern regions, respectively.

\subsubsection{Correlation between the shock velocity and X-ray spectrum}
\label{sec:shock_velocity_analysis}
The shock velocity of the SNR has only been estimated for its northern part \citep{Katsuda_2008, Allen_2015}.
For the purpose of comparison, we assumed that the distance between the shock and the location where the progenitor star exploded provides an estimate (a proxy) of the average shock velocity since the explosion.
Precise measurements of the location where the progenitor star exploded are essential for accurately estimating shock velocity, although this point remains a topic of debate.
From Chandra observation, \citet{Mignani_2019} estimated the proper motion of the CCO as $\lesssim 300\,{\rm mas\,yr^{-1}}$ ($3\sigma$ upper limit).
In this case, the position at which the progenitor star exploded will be within the radius of $\lesssim 15\,{\rm arcmin}$ from the current position of the CCO, assuming that SN occurred 3,000 years ago.
Conversely, by observing the optical/infrared counterpart candidate of the CCO \citep{Mignani_2007} using the VLT, \citet{Mignani_2019} estimated the proper motion of the CCO as $\lesssim 10\,{\rm mas\,yr^{-1}}$ ($3\sigma$ upper limit).
If this value is correct, the explosion site of the progenitor star would be within the radius of $\lesssim 30''$ from the current position of the CCO, again assuming that SN occurred 3,000 years ago.
It remains unclear which estimate is correct. 
However, for the latter measurement, a value derived from an image fitting only the northern region of the SNR with a circular model $(14.7\pm6.6\, {\rm \,mas\,yr^{-1}})$ was consistent within the margin of error \citep{Camilloni_2023}.
Based on the above result, we assumed that the progenitor caused the SN at the current location of the CCO (R.A\,=\,\timeform{08h52m1s.4}, Dec.\,=\,$-$\timeform{46D17'53''.3} (J2000.0)) \citep{Pavlov_2001}, and considered the distances between the shock and the CCO to be the average shock velocity since explosion.

\begin{figure*}
\begin{center}
\includegraphics[width=15cm,clip]{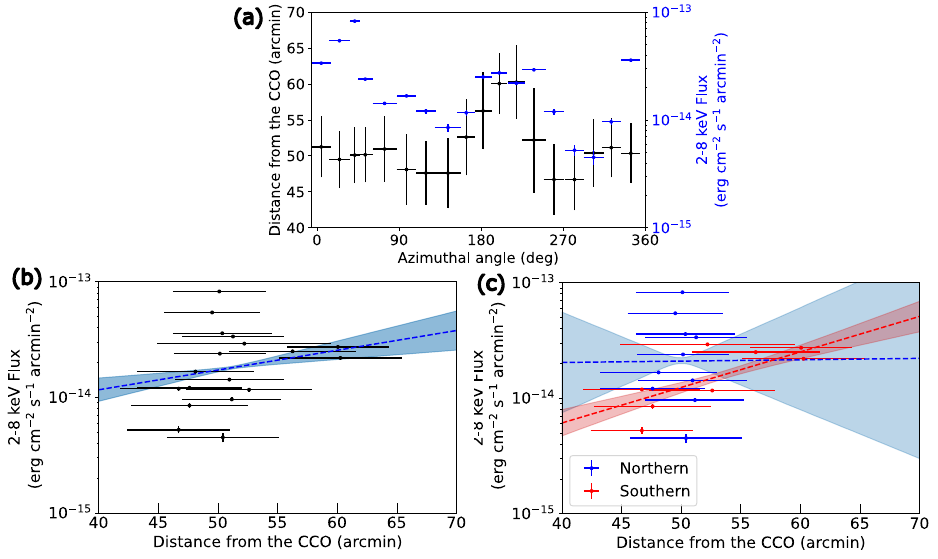}
\end{center}
\caption{
        Relation between X-ray flux and shock velocity.
        Figure (a) represents the azimuthal distribution of distances between shock front and the CCO ($\approx$ shock velocity) (black) and X-ray flux (blue).
        Figure (b) represents the correlation plot with best-fit results of a linear model ($\log F_{\rm X} = a (D_{\rm CCO}-c) + b$) (blue dash line) and 1$\sigma$ error (blue filled area).
        Figure (c) shows the correlation plot of the northern and southern regions, represented by blue and red lines, respectively.
        We note that errors in figure (b) and (c) are $1\sigma$ confidence level.
        {Alt text: Three graphs.
        In figure (a), the x-axis shows the azimuthal angle of minus 6 to 360 degrees.
        The y-axis on the left shows the distance from the CCO from 40 to 70 arc-minutes.
        The y-axis on the right shows the 2 to 8 kilo-electron volt X-ray flux from 10 to the power of minus 15 to 10 to the power of minus 13.
        In figures (b) and (c), the x-axis shows the distance from the CCO from 40 to 70 arc-minutes.
        The y-axis shows the 2 to 8 kilo-electron volt X-ray flux from 10 to the power of minus 15 to 10 to the power of minus 12.}
}
\label{fig:Xflux_vs_distance}
\end{figure*}

The resultant azimuthal distribution of the CCO distance from the rim (proxy of the shock velocity) is shown in figure \ref{fig:Xflux_vs_distance} (a) together with the X-ray flux profile and plot of the change in the X-ray flux in response to the shock velocity for total (b) and northern and southern regions (c).
We fitted the correlation plot with a logarithmic model ($\log F_{\rm X} = a (D_{\rm CCO}-c) + b$, where $D_{\rm CCO}$ represents the distance from the CCO).
The best-fit parameters are shown in table \ref{tab:velajr_corelation}.
The significance of the correlation is $1.85\sigma$, $0.03\sigma$, and $3.80\sigma$ for total, northern, and southern regions, respectively.

\begin{figure*}
\begin{center}
\includegraphics[width=15cm,clip]{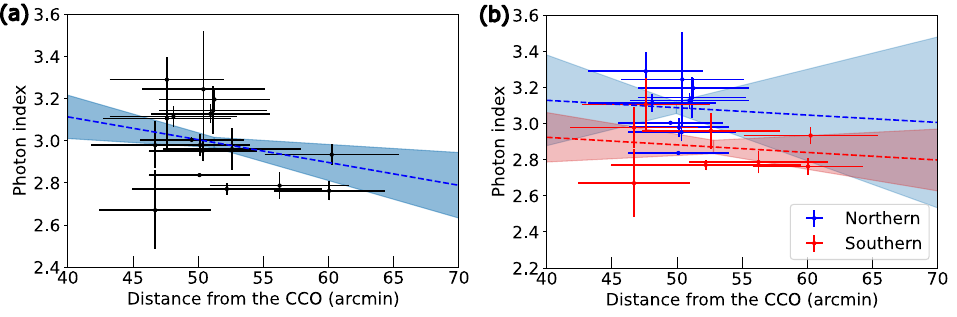}
\end{center}
\caption{
        Correlation plot of the photon index and the shock velocity.
        The blue dashed lines and blue filled area of figure (a) represent the best-fit results of a linear model ($\Gamma = a (D_{\rm CCO}-c) + b$) and 1$\sigma$ error, respectively.
        Figure (b) shows the correlation plot of the northern and southern regions, represented by blue and red lines, respectively.
        We note that errors in the right figure are $1\sigma$ confidence level.
        {Alt text: Two graphs.
        The x-axis shows the distance from the CCO from 40 to 65 arc-minutes.
        The y-axis shows the photon index from 2.4 to 3.6 and 2.2 to 3.6 for figure (a) and figure (b), respectively}.
}
\label{fig:index_vs_distance}
\end{figure*}

Figure \ref{fig:index_vs_distance} (a) and (b) shows a plot of the changes in the photon index in response to the shock velocity for the total and northern and southern regions, respectively.
We fitted the correlation plot with a linear model ($\Gamma = a (D_{\rm CCO}-c) + b$).
The best-fit parameters are shown in table \ref{tab:velajr_corelation}.
The significance of the anti-correlation is $1.55\sigma$, $0.18\sigma$, and $0.92\sigma$ for total, northern, and southern regions, respectively.

\begin{table*}
\tbl{Correlation coefficient between parameters and best-fit parameters\footnotemark[$*$]}{
\begin{tabular}{cccccccc}
\hline
  \multicolumn{3}{c}{Parameter} & Region & Significance of correlation & \multicolumn{3}{c}{Best-fit parameter} \\
  X-axis & & Y-axis & & $\sigma$ & $a \pm \Delta a$ & $b \pm \Delta b$ &  $c$ \\
\hline
X-ray flux       & vs. & Photon index   & Total     & 1.47       & $(-1.6 \pm 1.1) \times 10^{-1}$      & $2.99\pm0.02$      & $-13.743$                \\
                 &     &                & Northern  & 2.10       & $(-3.0 \pm 1.4) \times 10^{-1}$      & $3.09\pm0.03$      & $-13.680$                \\
                 &     &                & Southern  & 0.55       & $(-1.0 \pm 1.9) \times 10^{-1}$      & $2.87\pm0.04$      & $-13.822$                \\\hline
Cloud density    & vs. & X-ray flux     & Total     & 5.12       & $(1.3 \pm 0.2) \times 10^{-22}$      & $-13.743\pm 0.004$ & $3.23 \times 10^{21}$ \\ 
                 &     &                & Northern  & 4.10       & $(2.1 \pm 0.5) \times 10^{-22}$      & $-13.680\pm 0.005$ & $3.8 \times 10^{21}$ \\ 
                 &     &                & Southern  & 0.39       & $(0.5 \pm 1.4) \times 10^{-22}$      & $-13.822\pm 0.006$ & $2.5 \times 10^{21}$ \\\hline 
Cloud density    & vs. & Photon index   & Total     & 1.84       & $(5.1 \pm 2.8) \times 10^{-23}$      & $2.99\pm0.02$      & $3.23 \times 10^{21}$ \\
                 &     &                & Northern  & 1.17       & $(-7.6 \pm 6.5) \times 10^{-23}$     & $3.09\pm0.03$      & $3.8 \times 10^{21}$ \\
                 &     &                & Southern  & 0.54       & $(-0.6 \pm 1.1) \times 10^{-23}$     & $2.87\pm0.04$      & $2.5 \times 10^{21}$ \\\hline
Gamma-ray counts & vs. & X-ray flux     & Total     & 7.92       & $(4.2 \pm 0.5) \times 10^{-3}$       & $-13.743\pm 0.004$ & $201$                        \\
                 &     &                & Northern  & 6.62       & $(5.0 \pm 0.7) \times 10^{-3}$       & $-13.680\pm 0.005$ & $224$                        \\
                 &     &                & Southern  & 2.86       & $(3.7 \pm 1.3) \times 10^{-3}$       & $-13.822\pm 0.006$ & $172$                        \\\hline
Gamma-ray counts & vs. & Photon index   & Total     & 0.44       & $(2.7 \pm 6.1) \times 10^{-4}$       & $2.99\pm0.02$      & $201$                        \\
                 &     &                & Northern  & 1.18       & $(-1.1 \pm 1.0) \times 10^{-3}$      & $3.09\pm0.03$      & $224$                        \\
                 &     &                & Southern  & 0.62       & $(-8 \pm 13) \times 10^{-4}$         & $2.87\pm0.04$      & $172$                        \\\hline
Shock velocity   & vs. & X-ray flux     & Total     & 1.85       & $(1.7 \pm 0.9) \times 10^{-2}$       & $-13.743\pm0.004$  & $51.2$                     \\
                 &     &                & Northern  & 0.03       & $(1 \pm 46) \times 10^{-3}$          & $-13.680\pm0.005$  & $50.0$                     \\
                 &     &                & Southern  & 3.80       & $(3.1 \pm 0.8) \times 10^{-2}$       & $-13.822\pm0.006$  & $53$                     \\\hline
Shock velocity   & vs. & Photon index   & Total     & 1.55       & $(-1.1 \pm 0.7) \times 10^{-2}$      & $2.99\pm0.02$      & $51.2$                     \\
                 &     &                & Northern  & 0.18       & $(-4 \pm 22) \times 10^{-3}$         & $3.09\pm0.03$      & $49.9$                     \\
                 &     &                & Southern  & 0.56       & $(-4 \pm 8) \times 10^{-3}$          & $2.87\pm0.04$      & $53$                     \\\hline
Cloud density    & vs. & Shock velocity & Total     & 2.17       & $(-1.5 \pm 0.7) \times 10^{-21}$     & $51.2\pm0.6$       & $3.23 \times 10^{21}$ \\
                 &     &                & Northern  & 0.16       & $(-2 \pm 11) \times 10^{-21}$        & $50.0\pm0.8$       & $3.8 \times 10^{21}$ \\
                 &     &                & Southern  & 0.92       & $(-3 \pm 4) \times 10^{-21}$         & $53\pm1$           & $2.5 \times 10^{21}$ \\
\hline
\end{tabular}}\label{tab:velajr_corelation}  
\begin{tabnote}
\footnotemark[*] The errors represent a $1\sigma$ confidence level on an interesting single parameter.  \\ 
\end{tabnote}
\end{table*}

\section{Discussion}
\label{sec:discussion}

The X-ray spectrum obtained along the rim of RX J0852.0-4622 by Suzaku can be well explained by an absorbed power-law model with a photon index of $\sim 2.8$.
We also quantitatively evaluated the azimuthal variations of the X-rays according to flux and photon index.
The trends in the azimuthal distribution of flux and photon index are also consistent with the results obtained by SRG/eROSITA and XMM-Newton/EPIC \citep{Camilloni_2023}.
In this section, we discuss the origin of the azimuthal variations in synchrotron X-rays possibly due to (1) the angle between the shock and the Galactic magnetic field, variations in (2) the cloud density, and (3) shock velocity.

\subsection{Azimuthal variation by the Galactic magnetic field}
From magnetohydrodynamic analyses, the efficiency of charged particles injected into the shock region and the magnitude of the magnetic turbulence differs as the angle between the shock normal and the magnetic field changed (e.g., \cite{Leckband_1989, Fulbright_1990, Reynolds_1998, Amano_2007}).
These phenomena cause the bilateral distribution of X-rays and gamma-rays, as represented in the shell-type SNR SN 1006 \citep{Rothenflug_2004}.

\begin{figure}
    \begin{center}
    \includegraphics[width=8cm,clip]{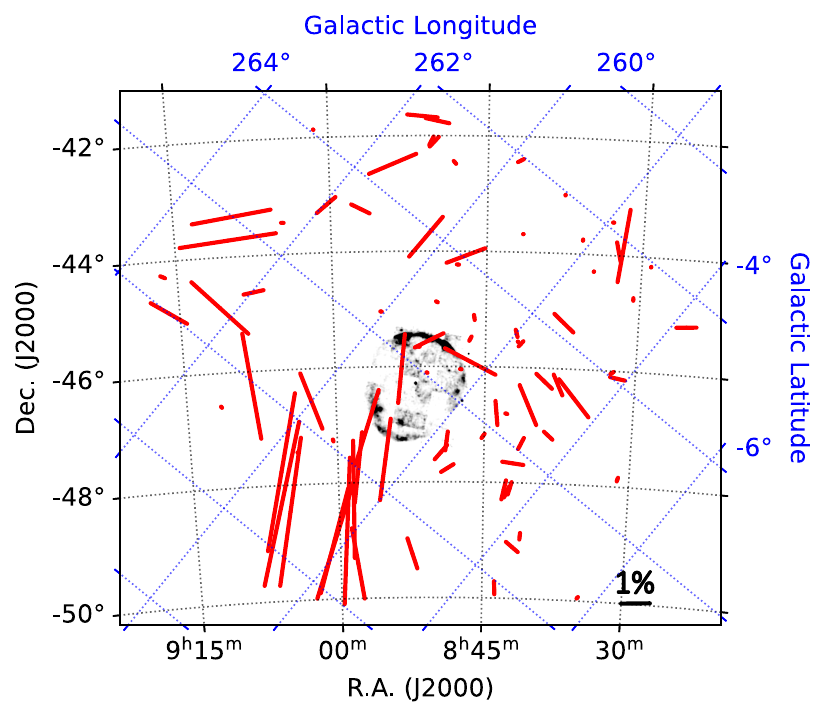}
    \end{center}
    \caption{
        The Galactic magnetic field around the RX~J0852.0$-$4622 estimated by the polarization of stars.
        The direction of the red lines represents the direction of the Galactic magnetic field.
        The length of the red line represents the polarization degree, which is proportional to the intensity of the Galactic magnetic field.
        The black line in the lower right of the figure indicates the length of a 1\% polarization degree.
        The background X-ray image is from Suzaku/XIS observations in the energy band 2.0-5.0 keV.
        {Alt text: A graph.
         X-ray image of SNR RX~J0852.0$-$4622 with the magnetic field estimated by the polarization of stars.}
    }
    \label{fig:velajr_mag}
\end{figure}

To investigate this scenario, we estimated the direction of the Galactic magnetic field around the SNR from the polarization of starlight in 0--2 kpc of Earth, as summarized in the stellar polarization catalog \citep{Heiles_2000}.
The estimated Galactic magnetic field around the SNR is shown in figure\ref{fig:velajr_mag}, where the direction of the red line represents the direction of the Galactic magnetic field.
The length of the red line represents the polarization degree, which is proportional to the intensity of the Galactic magnetic field.
As shown in figure \ref{fig:velajr_mag}, the Galactic magnetic field around the SNR was distributed unevenly.
Additionally, the fast stellar wind of a massive progenitor star creates magnetic turbulence in the circumstellar material \citep{Biermann_1993, Zirakashvili_2018b}.
These results lead us to conclude that the magnetic fields upstream of the shock are not aligned with the Galactic magnetic field.

Based on these facts, we conclude that the observed azimuthal variation in the X-ray spectrum are most likely not caused by the Galactic magnetic field.
To investigate the effect of the Galactic magnetic field on cosmic-ray acceleration in more detail, it is necessary to accurately determine its direction around the SNR and the magnetic field downstream of the shock.
The former may be possible from future stellar polarization observations such as SOUTH POL \citep{Magalha_2012} and others.
For the latter, observation of the polarization of X-ray synchrotron radiation by the Imaging X-ray Polarimetry Explorer (IXPE) \citep{Weisskopf_2022} will be an effective method.

\subsection{Azimuthal variation by cloud density}
As shown in sections \ref{sec:X-ray_spectrum_and_cloud_density} and \ref{sec:x-ray_vs_gamma-ray}, the X-ray and gamma-ray fluxes are positively related to the cloud density.
In this section, we interpret these results based on the shock-cloud interaction model.

\subsubsection{Shock-cloud interaction}
\label{sec:velajr_shock_cloud_inter}
Numerical magneto-hydrodynamic simulations indicate that the magnetic field and magnetic turbulence are amplified when the shock interacts with a dense cloud \citep{Inoue_2009,Inoue_2012b}.
Due to this magnetic field amplification, the synchrotron X-ray flux will positively correlate with the cloud density in the several-parsec scale.
This trend is qualitatively consistent with the global positive correlation between cloud density and X-ray flux obtained by our analysis.

As shown in the figure \ref{fig:flux_vs_nh} (b) and (c), the X-ray flux and cloud density correlate very well in the northern part of the SNR with a significance of $4.10\sigma$.
These areas have a relatively high cloud density, indicating a strong influence of the shock-cloud interaction.
In contrast to the northern part of the SNR, X-ray flux was also high in the southern part, where the cloud density was low.
Indeed, the significance of the correlation between X-ray flux and cloud density was low ($0.39\sigma$) in the southern region.
The shock-cloud interaction model cannot explain this phenomenon, and factors other than cloud density may be at play and is discussed in section \ref{sec:shock_velocity_discussion}.

The interaction between the shock and clouds is considered to have started within 1,000 years after the SN, and the magnetic amplification via the interaction occurred in the recent several 100 years \citep{Fukui_2024}.
In this case, the electrons near the shock would no longer emit X-rays due to cooling and would be emitted primarily from further downstream regions, where the magnetic field has not been amplified.
The electrons are then affected by the magnetic field of the region that they pass as they travel downstream, and therefore, some relation between the roll-off energy and cloud density is expected to be observed.
This expectation is inconsistent with the results; no significant correlation was found between the photon index and cloud density.
One possible reason for this is that the elapsed time from the interaction with the shock and cloud is different in each region.
To elucidate the cause, a detailed comparison between the spatial distribution of the X-ray spectrum and the cloud density is needed.
Observation by X-ray astronomy satellites with higher angular resolution is required in the future.

\subsubsection{Comparison with other SNRs}
There are similar studies that have compared the spatial variation of the X-ray spectrum and cloud densities of SNR RX~J1713.7$-$3946 \citep{Sano_2015, Fukui_2024}.
SNR RX~J1713.7$-$3946 is a core-collapsed SNR with an age of $\sim 1,600$ yrs old \citep{Tsuji_2016, Acero_2017} and features strong non-thermal emission \citep{Koyama_1997, Slane_1999, Uchiyama_2003, Cassam_2004, Tanaka_2008, Acero_2009, Higurashi_2020, Tanaka_2020}, similar to SNR RX~J0852.0$-$4622.
The X-ray flux of the supernova remnant RX~J1713.7$-$3946 is positively correlated with cloud density; however, the photon index does not show any correlation.
Additionally, there is a negative correlation between the photon index and the X-ray flux. 
These trends are similar to our findings for the supernova remnant RX~J0852.0$-$4622. 
\citet{Sano_2015} concluded that these correlations result from shock-cloud interactions, further supporting our shock-cloud interaction scenario.
Similar positive correlations between synchrotron X-ray flux and cloud density are also discovered in young SNRs such as the northern region of Kepler SNR \citep{Sapienza_2022} and the east region in Tycho SNR \citep{Lopez_2015}.

On the other hand, some SNRs showed an anti-correlation between cloud density and synchrotron X-ray flux: southwest \citep{Miceli_2014} and northwest region \citep{Bamba_2008, Sano_2022, Ichihashi_2024} of SN 1006, and RCW 86 \citep{Bamba_2023}.
\citet{Sano_2021} suggested that these differences are most likely caused by the difference in the radius of the clouds interacting with the shock.
Further investigation is required to elucidate the causes of these differences.

\subsection{Azimuthal variation by shock velocity}
\label{sec:shock_velocity_discussion}
\begin{figure*}
\begin{center}
\includegraphics[width=15cm,clip]{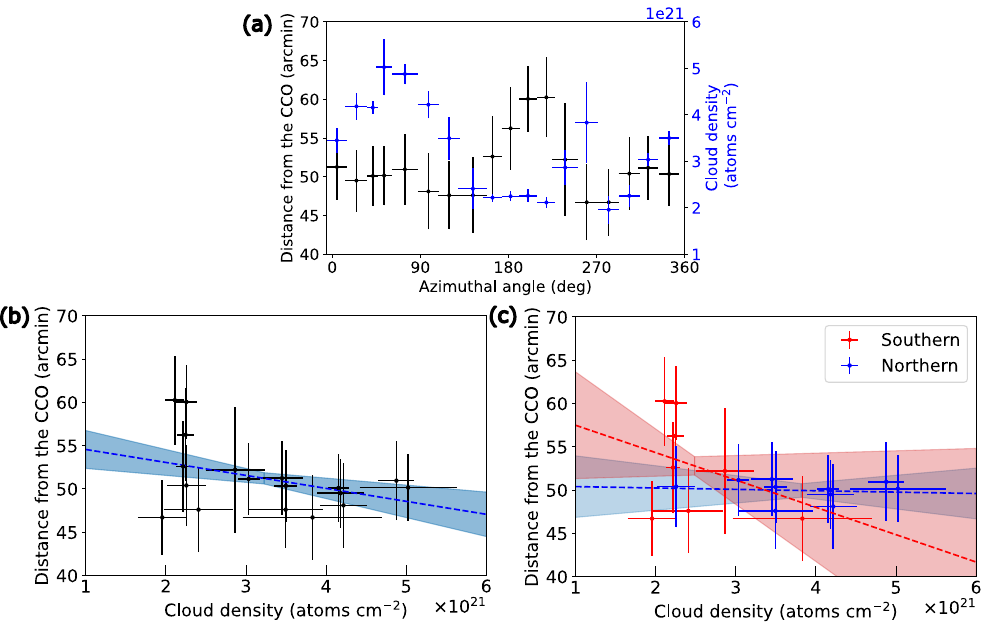}
\end{center}
\caption{
        Relation between shock velocity and cloud density.
        Figure (a) represents the azimuthal distribution of distances between shock front and the CCO ($\approx$ shock velocity) (black) and cloud density (blue).
        Figure (b) represents the correlation plot with best-fit results of a linear model ($D_{\rm CCO} = a (N_{\rm cloud}-c) + b$) (blue dash line) and 1$\sigma$ error (blue filled area).
        Figure (c) shows the correlation plot of the northern and southern regions, represented by blue and red lines, respectively.
        We note that errors in figures are $1\sigma$ confidence level.
        {Alt text: Tow graph.
        In figure (a), the x-axis shows the azimuthal angle of minus 6 to 360 degrees.
        The y-axis on the left shows the distance from the CCO from 40 to 70 arc-minutes.
        The y-axis on the right shows the cloud density from 1 times 10 to the power of 21 to 6 times 10 to the power of 21 atoms per square centimeter.
        In figures (b) and (c), the x-axis shows the cloud density from 1 times 10 to the power of 21 to 6 times 10 to the power of 21 atoms per square centimeter.
        The y-axis shows the distance from the CCO from 40 to 70 arc-minutes.}
}
\label{fig:cloud_vs_distance}
\end{figure*}

The analysis in section \ref{sec:shock_velocity_analysis} shows that the average shock velocity of the SNR was constant in the northern part of the SNR.
In contrast, the southern and southeastern parts of the SNR ($\sim150^{\circ} - 270^{\circ}$) exhibited an average shock velocity that was roughly 1.5 times faster than that observed in other areas of the SNR.
This trend can be qualitatively explained by the lower cloud density.
To evaluate this scenario qualitatively, we assessed the correlation between shock velocity and cloud density.
Figure \ref{fig:cloud_vs_distance} (a), (b), and (c) shows the azimuthal distribution along the SNR rim of the shock velocity and cloud density, the plot of the changes in the shock velocity in response to the cloud density for total, and northern and southern regions, respectively.
This figure suggests that the averaged shock velocity and cloud density are anti-correlated.
We fitted the data with a linear model ($D_{\rm CCO} = a (N_{\rm cloud}-c) + b$) and plotted the best-fit result.
The significance of the anti-correlation is $2.17\sigma$, $0.16\sigma$, and $0.92\sigma$ for total, northern, and southern regions, respectively.
The best-fit parameters ($a, b, c$) are presented in table \ref{tab:velajr_corelation}.
This result shows a weak anti-correlation between cloud density and shock velocity of the whole and southern region of the SNR, albeit with low significance.

In the diffusive shock acceleration scheme, when the maximum energy of the electrons is determined by the equilibrium between synchrotron energy loss and acceleration, the roll-off energy of synchrotron photons $E_{\rm roll}$ is given as follows \citep{Aharonian_1999, Yamazaki_2006, Zirakashvili_2007}: 
\begin{equation}
    E_{\rm roll} \propto V_{\rm sh}^{2}
    \label{eq:rolloff_energy}
\end{equation}
Equation (\ref{eq:rolloff_energy}) indicates that the roll-off energy is larger (i.e., the photon index is smaller) in the region of higher shock velocity.
This result is qualitatively consistent with the anti-correlation between the shock velocity and photon index in the southern and southeastern regions of the SNR (see figure \ref{fig:index_vs_distance}).

Additionally, the X-ray flux is positively related to the roll-off energy.
This relation indicates that higher X-ray fluxes are expected in the south of the SNR, where the shock velocity is large.
This expectation is consistent with our results that the X-ray flux correlates with shock velocity in the southern regions (see figure \ref{fig:Xflux_vs_distance} (c)) with a significance of $3.80\sigma$.

In conclusion, while the global azimuthal variation of the X-ray spectrum remains unexplained, the variation in shock velocity can qualitatively account for the changes observed in the southern part of the SNR.
We note that this discussion assumes that the CCO has not moved much since the SN, and that the discussion remains open.

\subsection{Fraction of hadronic component to observed gamma-rays}
\begin{longtable}{ccccc}
  \caption{Trends in cloud density, X-ray flux and gamma-ray flux for each region}\label{tab:all_azimuthal_distribution}  
\hline\noalign{\vskip3pt} 
  Region & Angle (deg.) & Cloud density & X-ray flux & Gamma-ray flux \\   [2pt] 
\hline\noalign{\vskip3pt} 
\endfirsthead      
\hline\noalign{\vskip3pt} 
  Region & Angle (deg.) & Cloud density & X-ray flux & Gamma-ray flux \\   [2pt] 
\hline\noalign{\vskip3pt} 
\endhead
\hline\noalign{\vskip3pt} 
\endfoot
\hline\noalign{\vskip3pt} 
\multicolumn{2}{@{}l@{}}{\hbox to0pt{\parbox{160mm}{\footnotesize
\hangindent6pt\noindent
\hbox to6pt{\hss}\unskip%
}\hss}} 
\endlastfoot 
A & $\sim 330 - \sim 50 $  & High & High &  High\\
B & $\sim 50 - \sim 130 $  & High & Low  &  High\\
C & $\sim 130 - \sim 220 $ & Low  & High &  Low \\
D & $\sim 220 - \sim 270 $ & Low  & High &  High (PWN may contaminated)\\
E & $\sim 270 - \sim 330 $ & Low  & Low  &  Low \\
\end{longtable}
As shown in figure \ref{fig:X-ray_vs_gamma-ray}, gamma-ray and X-ray fluxes were well correlated throughout the SNR.
However, the correlation was not uniform across regions.
To evaluate this correlation, we compared the azimuthal distribution of X-ray flux, gamma-ray flux, and cloud density.
Based on the trends in these azimuthal distributions, we divided them into five more homogeneous rim sectors (A, B, C, D, E) as shown in figure \ref{fig:velajr_extract_reg}, \ref{fig:flux_vs_nh} (a), and \ref{fig:X-ray_vs_gamma-ray} (a) and described in table \ref{tab:all_azimuthal_distribution}.

In table \ref{tab:all_azimuthal_distribution}, we categorize the values of each parameter as high or low in relation to the mean value.
The cloud density, X-ray flux, and gamma-ray flux in region A were all higher than average.
Considering that the flux of hadronic and leptonic origin gamma-rays are proportional to the cloud density and X-ray flux, respectively, we conclude that the combination of leptonic and hadronic components can explain the observed gamma-ray.
In region B, the cloud density and gamma-ray flux were higher, and X-ray flux was lower than in the other regions.
These results suggest that the observed gamma-rays mainly originated from hadronic components.
In contrast to region B, cloud density and gamma-ray flux were low, and X-ray flux was higher in region C than in other regions, leading us to conclude that the leptonic components are the dominant component of the observed gamma-rays in region C.
In region E, the cloud density, X-ray flux, and gamma-ray flux were all lower than average, showing that both leptonic and hadronic models can explain the gamma-rays.
This trend is qualitatively consistent with the results of \citet{Fukui_2024}, which estimated the spatial distribution of the ratio of the hadronic component to the observed gamma-ray flux by comparing the spatial distribution of X-ray flux, gamma-ray flux, and cloud density.
We ignored the discussion of region D because the gamma-ray flux may be contaminated by the emission from nearby PWN PSR J0855$-$4644.

\subsection{Possible cause of the azimuthal variation of X-ray spectrum}
Based on the discussions above, the spatial variation of the X-ray spectrum is most likely primarily influenced by two factors: the amplification of the magnetic field due to the shock-cloud interaction in the northern part of the SNR, and the higher shock velocity in the southern part.
Because the shock velocity is also related to cloud density, the cloud density interacting with the shock may significantly contribute to the spatial variation of the X-ray spectrum.
On the other hand, the Galactic magnetic field scenario cannot be completely ruled out, given the current accuracy of Galactic magnetic field estimation.

\section{Summary and Conclusion}
We analyzed the X-ray data obtained by Suzaku/XIS to quantitatively evaluate the azimuthal variation of the X-ray spectrum along the rim of SNR RX~J0852.0$-$4622. 
The obtained results are summarized below. 
\begin{enumerate}
    \item 
        We found that X-rays from the shock region in the energy range of 2-8 keV exceed several times that of the X-ray background (CXB, GRXE, SWCX, and/or LHB).
        The excess X-ray energy spectrum was well described by the absorbed power-law model with an average photon index of about 2.8, indicating that the X-rays were from synchrotron radiation from accelerated electrons with energies above the cutoff. 
        The X-ray flux and photon index were distributed bimodally in relation to the azimuthal angle.      
    \item 
        Polarization observations of stars near the SNR indicated that the Galactic magnetic field around the SNR was unevenly distributed.       
        These results lead us to conclude that the Galactic magnetic field orientation may not be the main contributor to the observed azimuthal variation in the X-ray spectrum.       
        Further investigation of the Galactic magnetic field is required to evaluate this scenario.      
    \item 
        The X-ray flux was positively correlated with the cloud density, similar to the positive correlation between the gamma-ray flux and cloud density \citep{Fukui_2017}.       
        This result can be explained qualitatively by the amplification of the magnetic field by the shock-cloud interaction.       
        On the other hand, we could not find a significative correlation between the photon index and cloud density, which is presumably due to relatively large uncertainties and dispersion in the photon index measurements, and to the fact that the timescale on which the shock-cloud interaction occurred varied by several hundred years.     
    \item 
        In the southern part of the SNR, the X-ray flux was positively correlated with the average shock velocity (i.e. distance between the shock and the CCO).
        This result is qualitatively consistent with the magnetic field amplification by shock in the diffusive shock acceleration scheme.       
    \item 
        We conclude that the spatial variation of the X-ray spectrum is most likely influenced by two factors: the amplification of the magnetic field due to the shock-cloud interaction in the northern part of the SNR and a higher shock velocity in the southern part.
        Because the shock velocity is also related to cloud density, we conclude that the cloud density interacting with the shock may significantly contribute to the spatial variation of the X-ray spectrum.
\end{enumerate}

\begin{ack}
We would like to thank Dr. Makoto S. Tashiro, Dr. Kosuke Sato, Dr. Jacco Vink, Dr. Koji Kawabata, Dr. Yuji Sunada, Dr. Kouichi Hagino, and Dr. Eric Miller for their constructive comments.
We would like to deeply appreciate the anonymous reviewer for their constructive comments on improving this paper.
\end{ack}

\section*{Funding}
This work was supported by JSPS KAKENHI Grant Numbers 23KJ0296 (DT), 23H01211 (AB), 23K20862 (SK).

\section*{Data availability} 
The Suzaku data underlying this article are available in Data ARchives and Transmission System (DARTS), at https://data.darts.isas.jaxa.jp/pub/suzaku/

\appendix 
\section{Estimation of the foreground hydrogen column density}
\label{sec:AppendixB}
\begin{figure*}
\begin{center}
\includegraphics[width=\linewidth,clip]{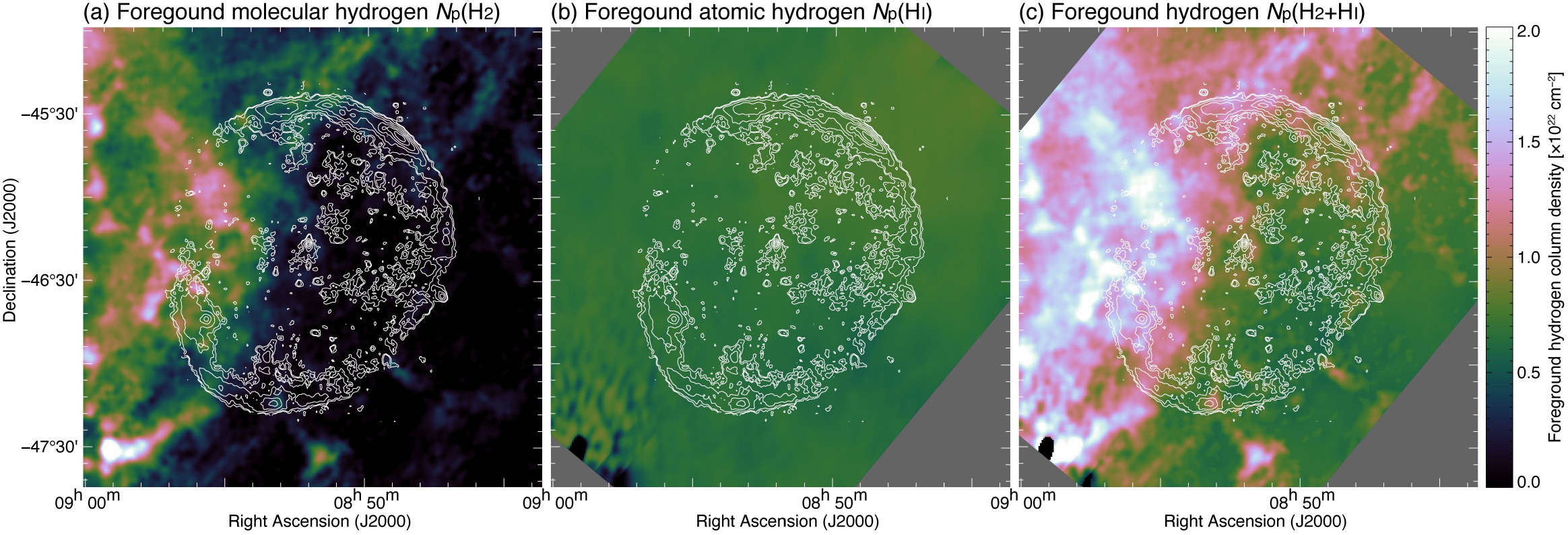}
\end{center}
\caption{Distributions of foreground hydrogen column density, $N_\mathrm{p}$, are shown. The integration velocity range of each map is from $-20$ to $+30$ km s$^{-1}$, corresponding to the foreground component for the SNR RX J0852.0$-$4622.
(a) $N_\mathrm{p}$(H$_2$) estimated from $^{12}$CO($J$~=~1--0). 
(b) $N_\mathrm{p}$(H{\sc i}) estimated from H{\sc i} by considering the H{\sc i} optical depth effect based on the Planck sub-millimeter dust optical depth. 
(c) $N_\mathrm{p}$(H$_2$ + H{\sc i}) estimated by summing up of $N_\mathrm{p}$(H$_2$) and $N_\mathrm{p}$(H{\sc i}). 
The superposed contours indicate the Suzaku X-ray intensity.
{Alt text: 3 cloud density maps with X-ray contours.}}
\label{fig:foreground_hydrogen}
\end{figure*}

In order to fix the absorbing column density of X-ray spectral modeling for each rectangular region, we estimated the foreground hydrogen column density for the SNR RX~J0852.0$-$4622. 
Following the method by \citet{Fukui_2024}, we estimated hydrogen column densities for molecular form $N_\mathrm{p}$(H$_2$) and atomic form $N_\mathrm{p}$(H{\sc i}) mainly using the NANTEN $^{12}$CO($J$~=~1--0) and ATCA \& Parkes H{\sc i} data. 
The velocity range of $-20$--$+30$~km~s$^{-1}$ was adopted by considering (1) the velocity range of shock-interacting clouds in RX~J0852.0$-$4622 ($V_\mathrm{LSR}$: $+20$--$+40$~km~s$^{-1}$) and (2) its expanding gas motion \citep{Fukui_2024}. 
We used the CO-to-H$_2$ conversion factor of $1.5 \times 10^{20}$~(K~km~s$^{-1}$)$^{-1}$ \citep{Aruga_2022}. 
The H{\sc i} column density $N_\mathrm{p}$(H{\sc i}) was corrected by considering the H{\sc i} absorption effect based on the Planck sub-millimeter dust optical depth following \citet{Fukui_2017}. 
Figure~\ref{fig:foreground_hydrogen} shows the distributions of foreground hydrogen column density, $N_\mathrm{p}$(H$_2$), $N_\mathrm{p}$(H{\sc i}), and $N_\mathrm{p}$(H$_2$+H{\sc i}). 
The spatial distribution of $N_\mathrm{p}$(H$_2$) and $N_\mathrm{p}$(H{\sc i}) were differed.
The spatial distribution of $N_\mathrm{p}$(H{\sc i}) are consistent throughout the SNR, and the western half of the SNR is dominated by $N_\mathrm{p}$(H{\sc i}) (typical values of which are $\sim$$7 \times 10^{21}$~cm$^{-2}$). 
By contrast, the eastern half of the SNR is dominated by $N_\mathrm{p}$(H$_2$), typical values of which are $\sim$1--$2 \times 10^{22}$~cm$^{-2}$.

\section{Best-fit results for X-ray spectra in each region}
\label{sec:X-ray_spectrum_list}
\begin{figure*}
\begin{center}
\includegraphics[width=0.3\textwidth,clip]{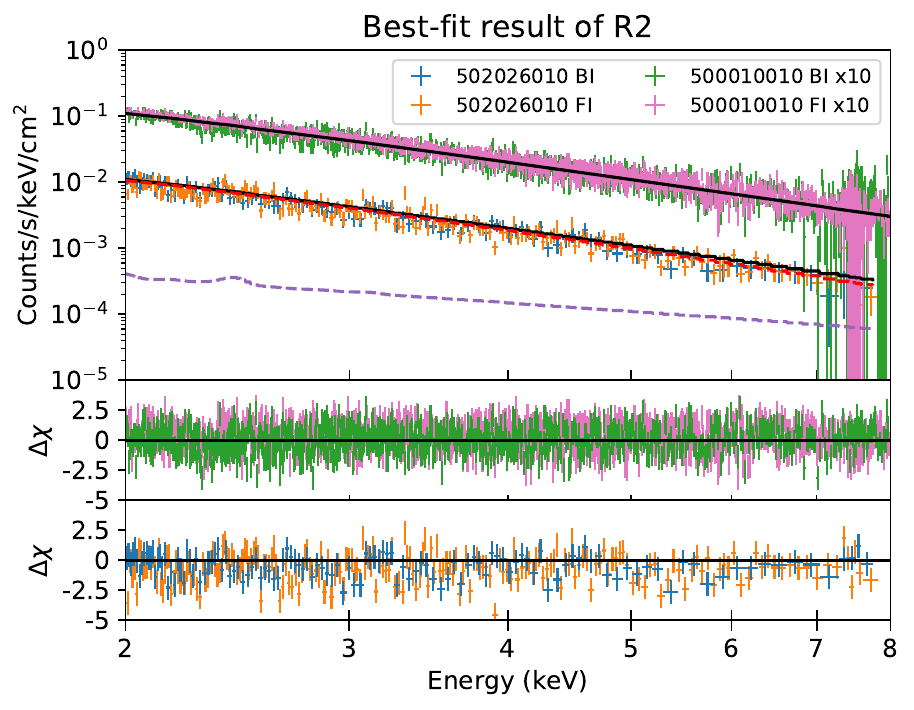}
\includegraphics[width=0.3\textwidth,clip]{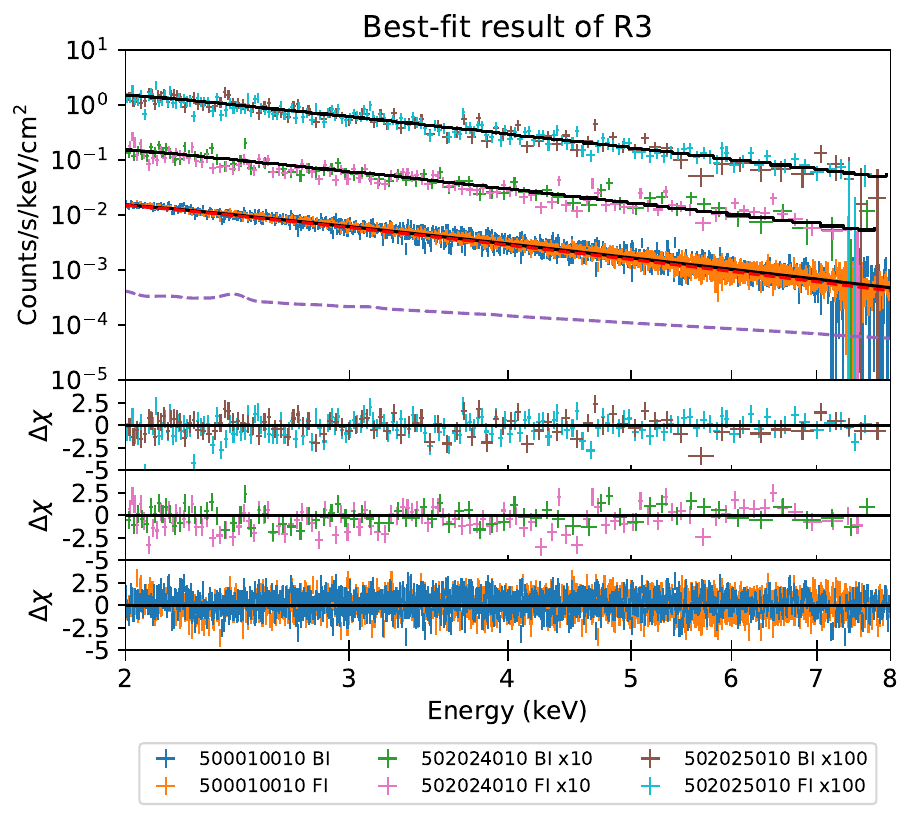}
\includegraphics[width=0.3\textwidth,clip]{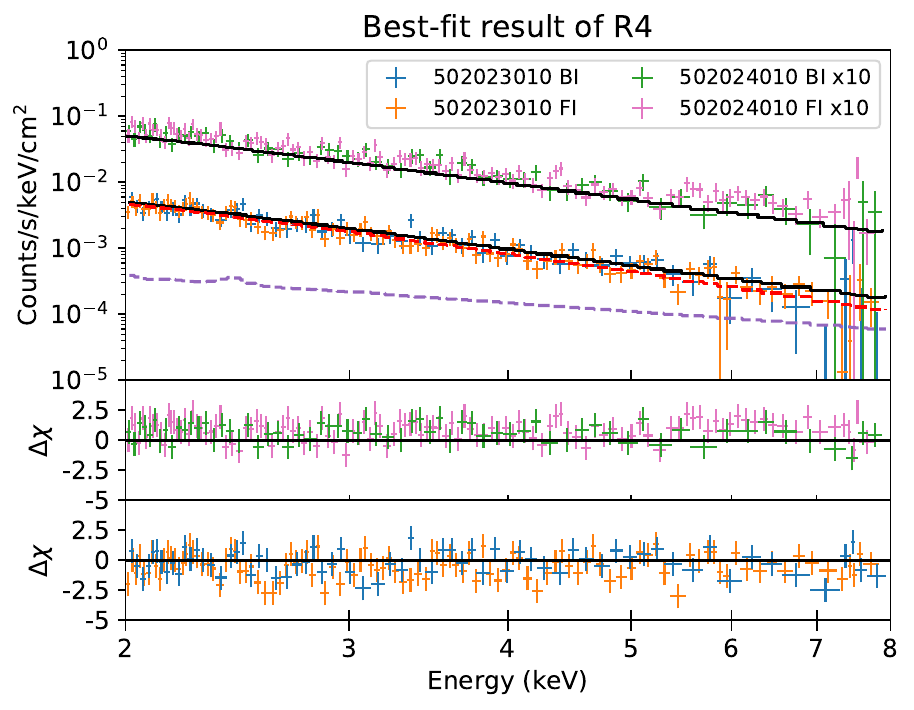}
\includegraphics[width=0.3\textwidth,clip]{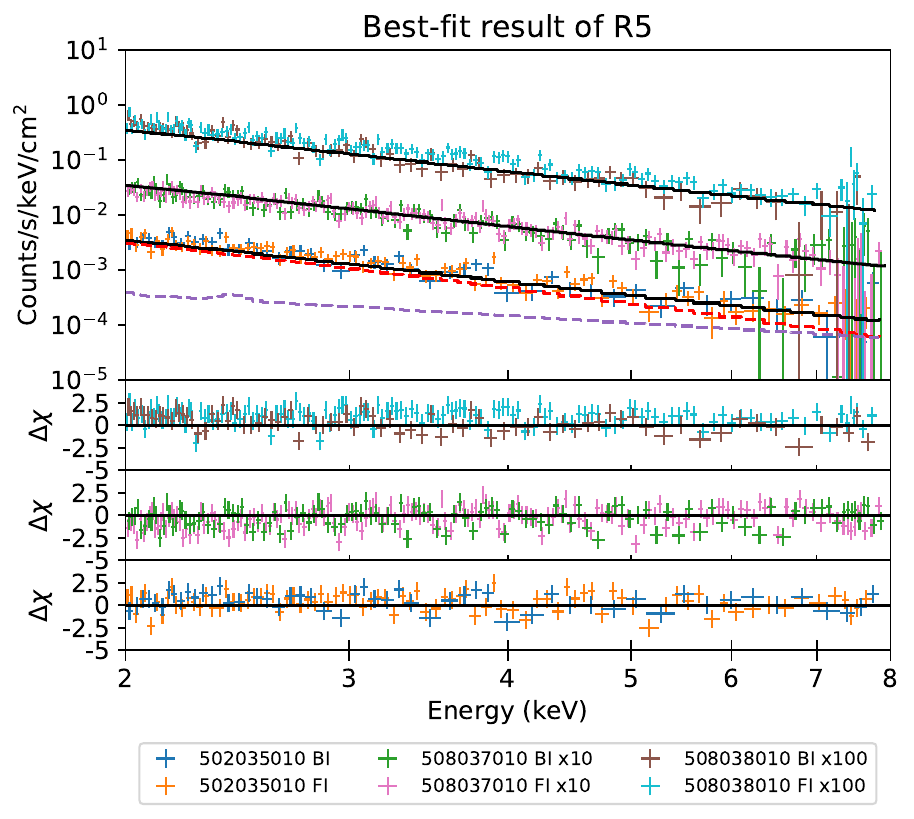}
\includegraphics[width=0.3\textwidth,clip]{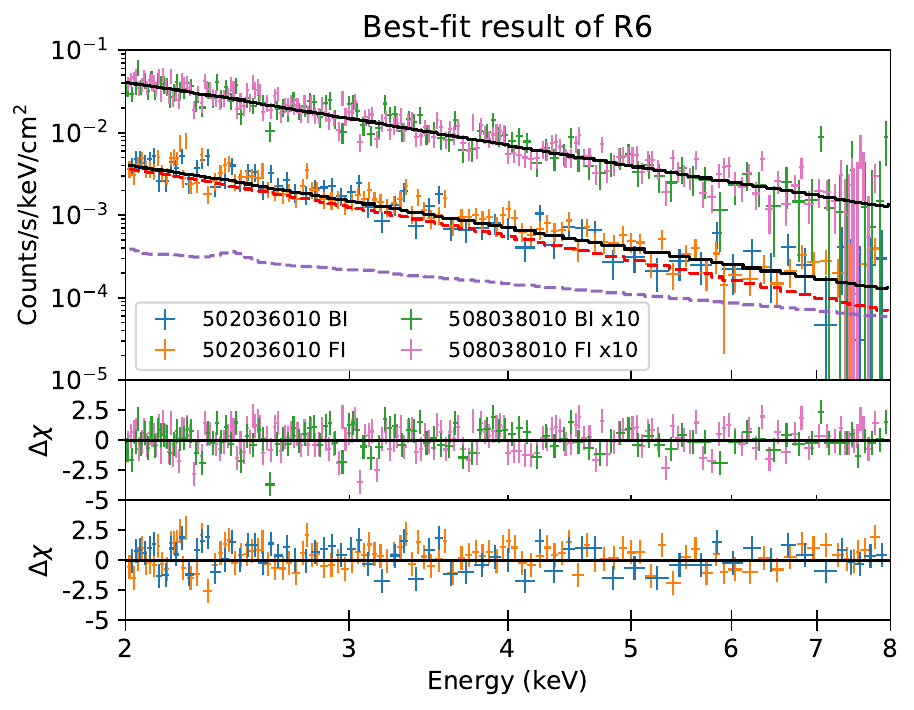}
\includegraphics[width=0.3\textwidth,clip]{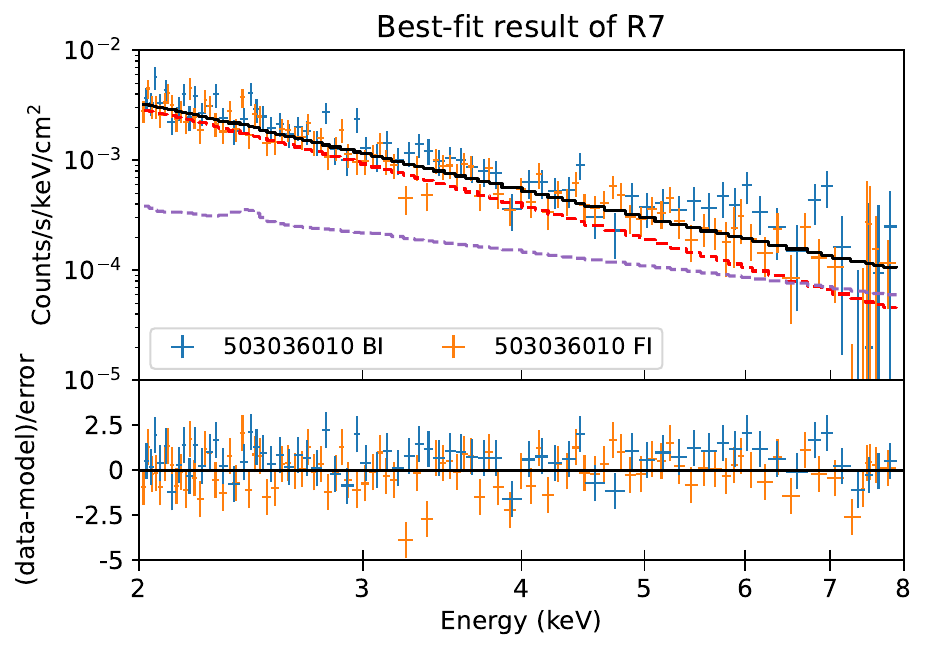}
\includegraphics[width=0.3\textwidth,clip]{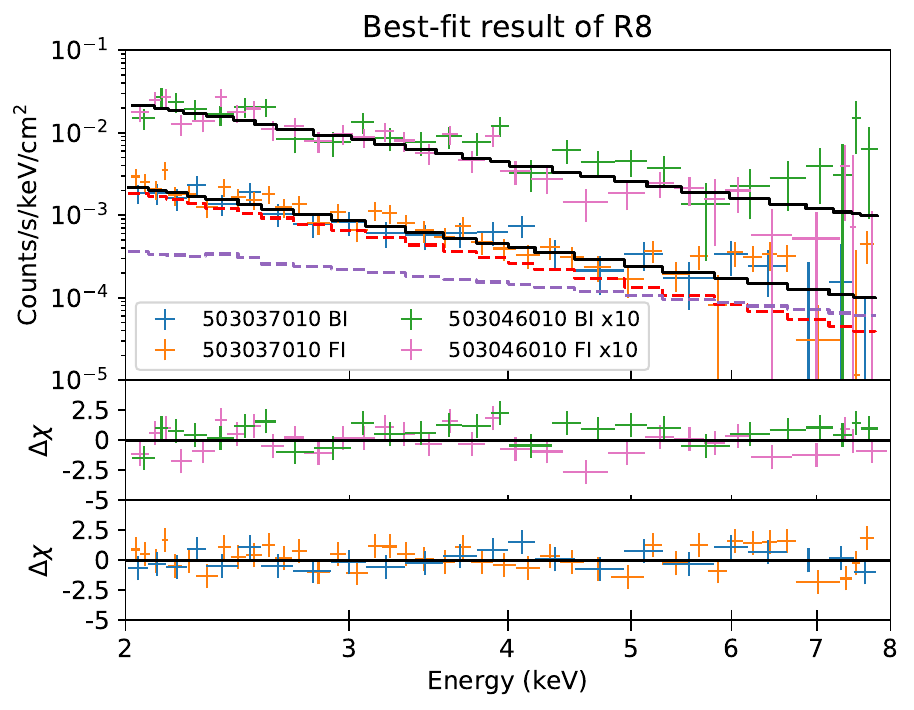}
\includegraphics[width=0.3\textwidth,clip]{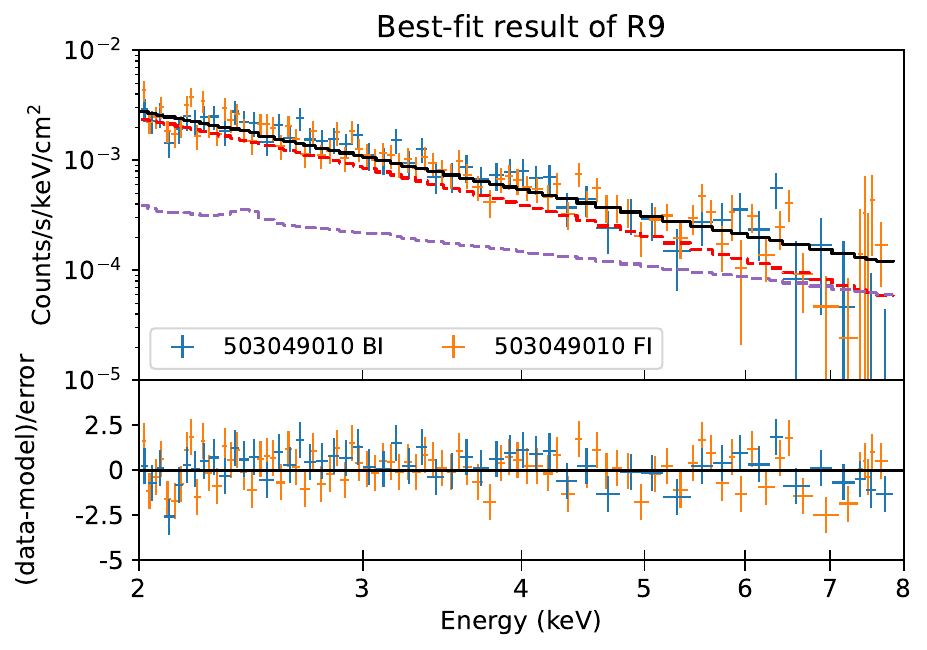}
\includegraphics[width=0.3\textwidth,clip]{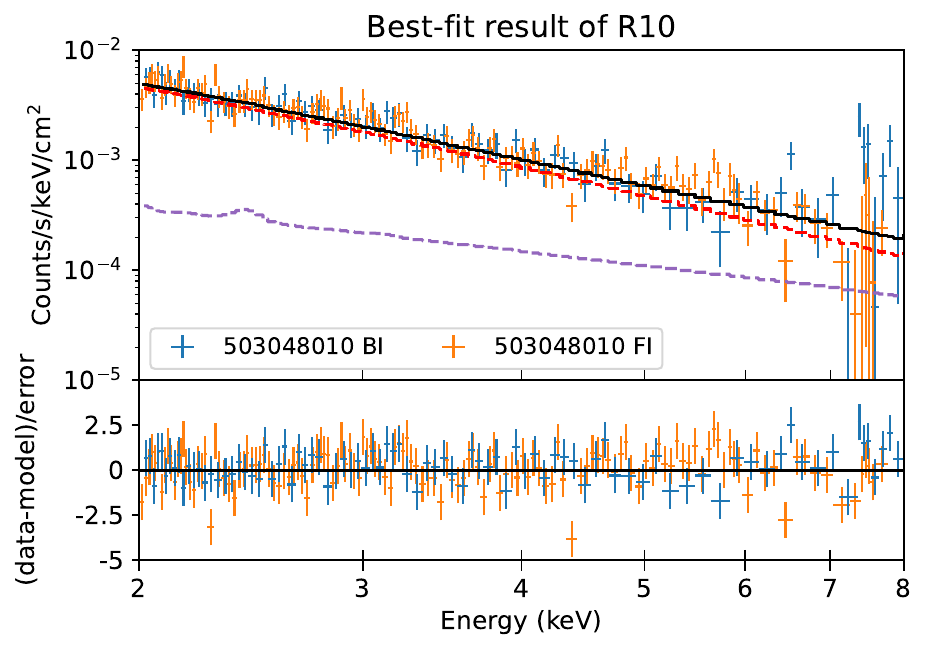}
\includegraphics[width=0.3\textwidth,clip]{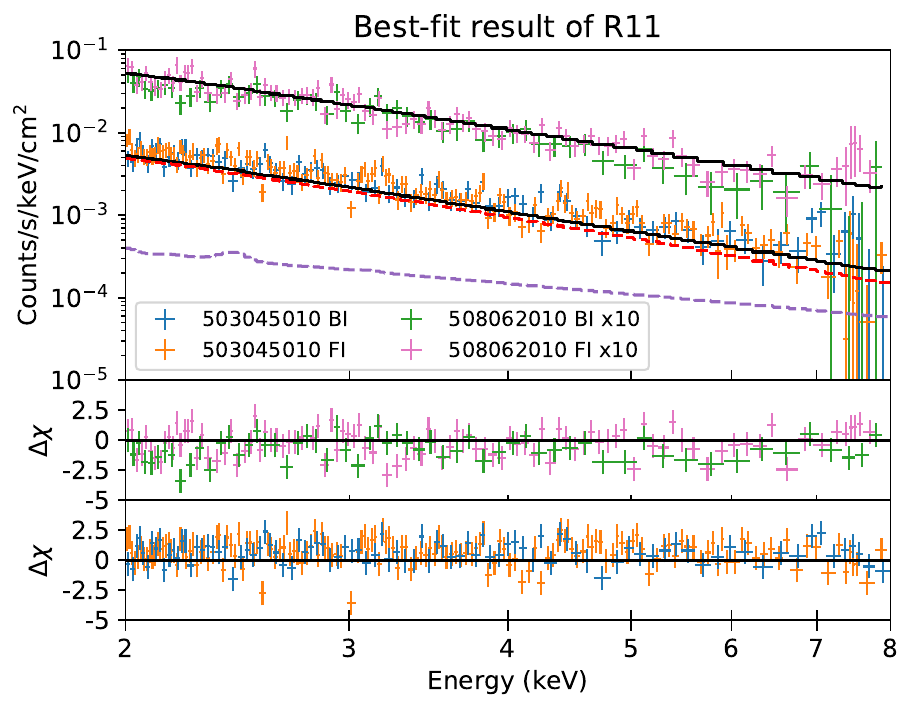}
\includegraphics[width=0.3\textwidth,clip]{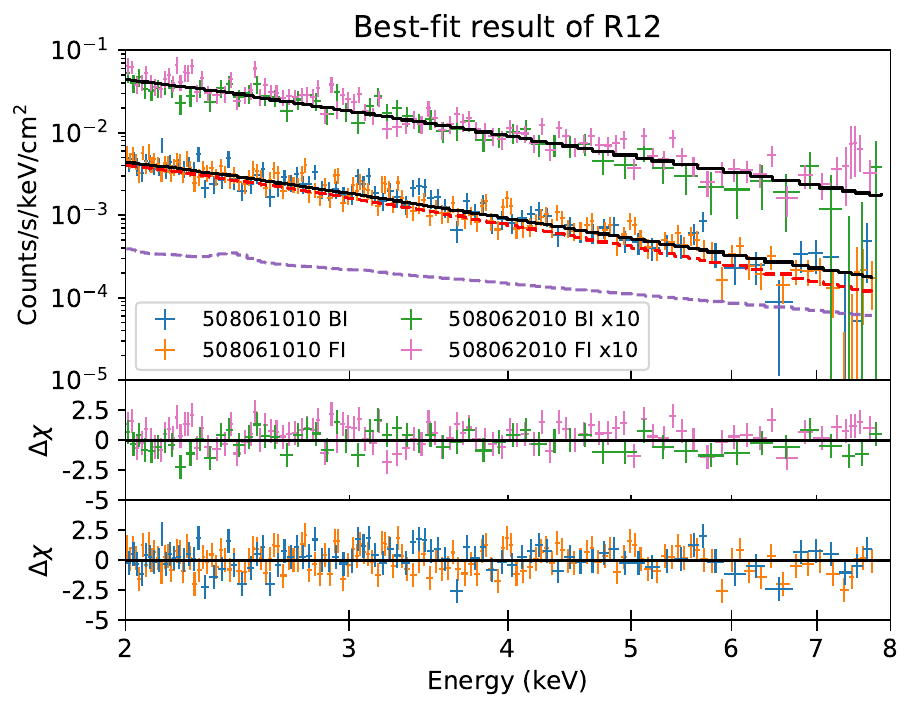}
\includegraphics[width=0.3\textwidth,clip]{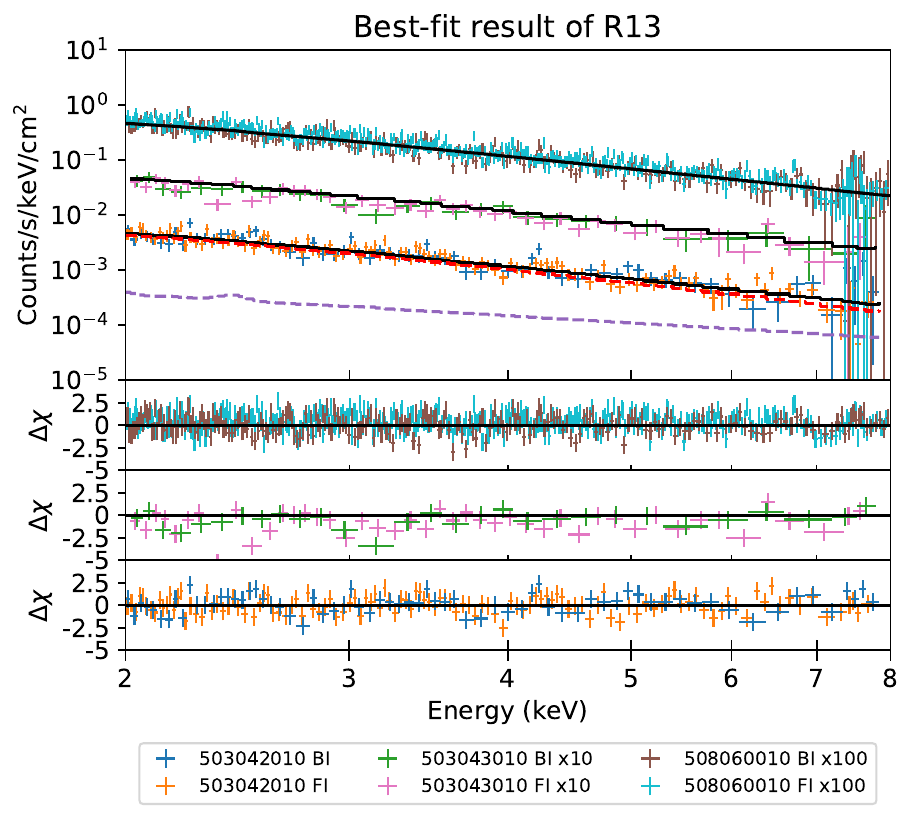}
\includegraphics[width=0.3\textwidth,clip]{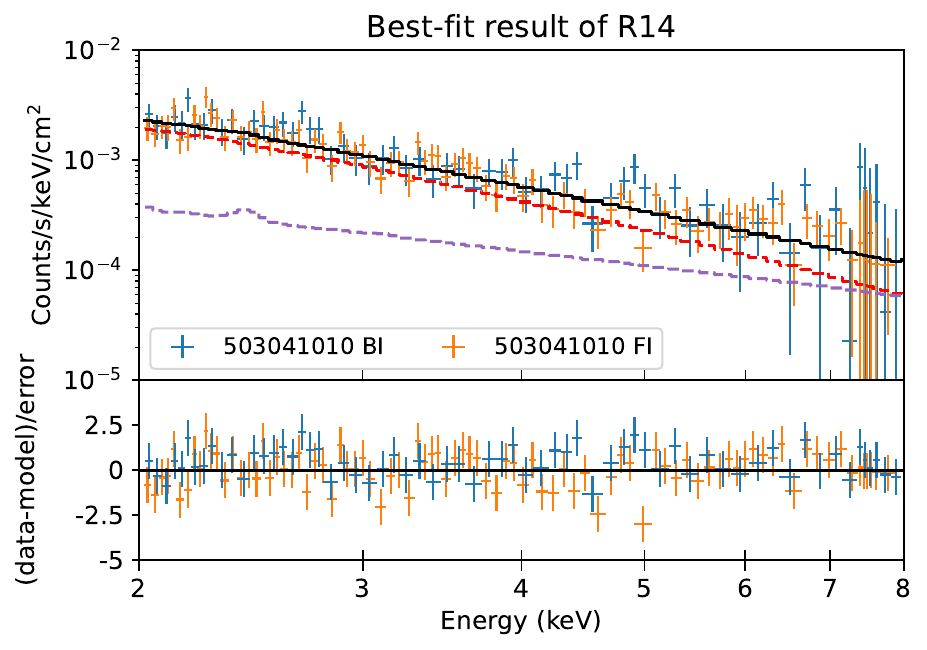}
\includegraphics[width=0.3\textwidth,clip]{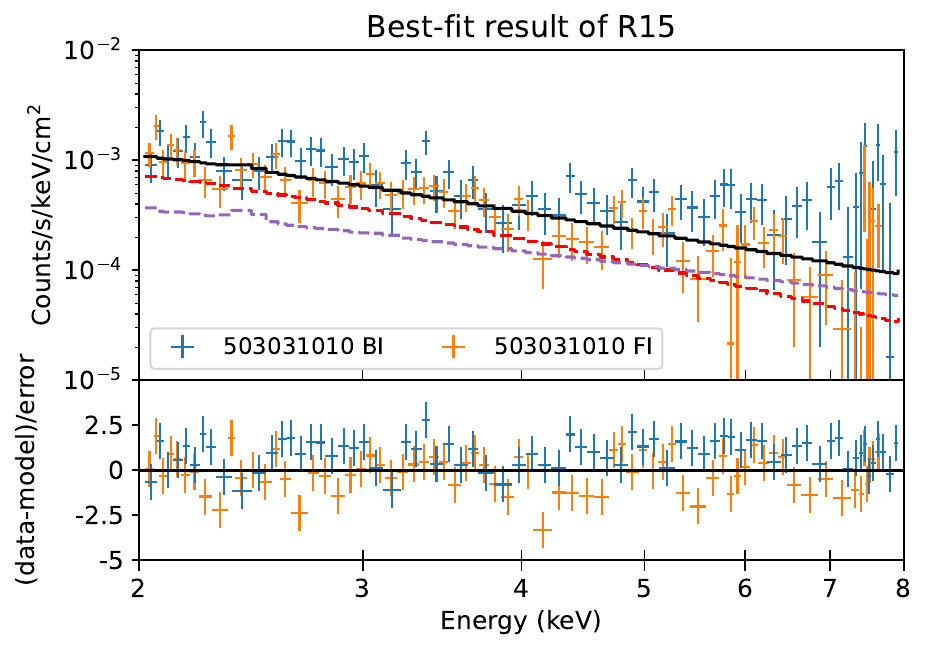}
\includegraphics[width=0.3\textwidth,clip]{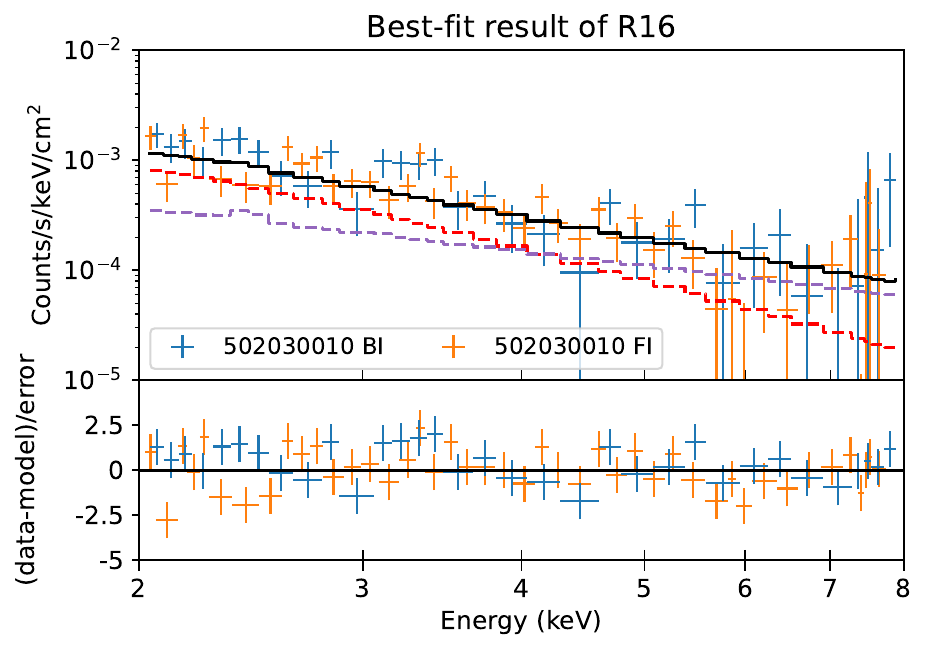}
\end{center}
\caption{
    The best-fit results of the NXB-subtracted X-ray spectrum for regions R2 to R16.
    The purple and red dashed lines represent the X-ray background (GRXE + CXB + SWCX) and absorbed power-law components, respectively.
    The black lines represent the total best-fit model.
    {Alt text: 15 line graphs.}
}
\label{fig:appendix_velajr_2to8keV_1}
\end{figure*}

\begin{figure*}
\begin{center}
\includegraphics[width=0.3\textwidth,clip]{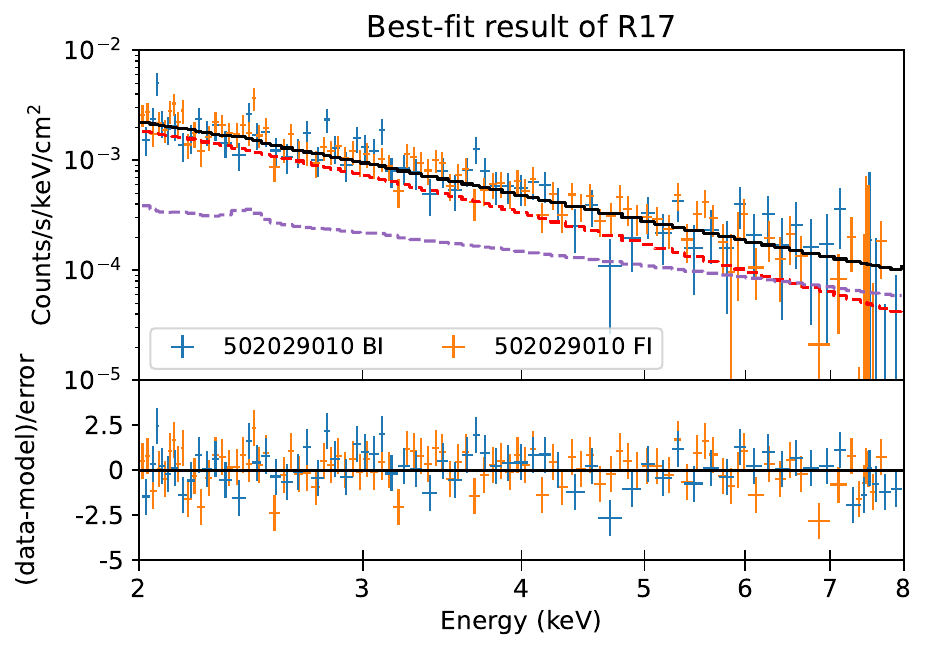}
\includegraphics[width=0.3\textwidth,clip]{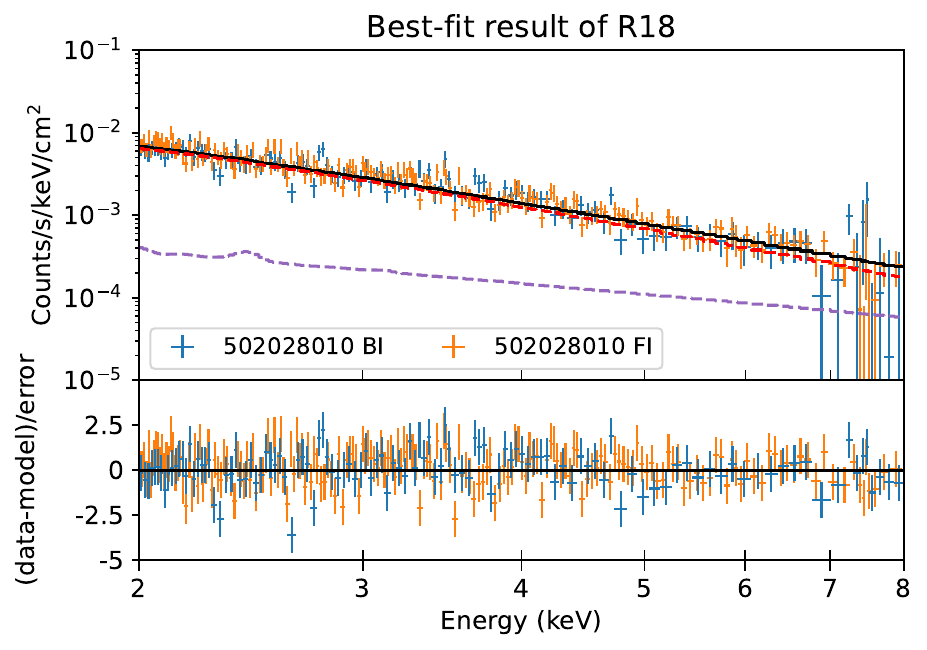}
\end{center}
\caption{As for figure \ref{fig:appendix_velajr_2to8keV_1}, but for regions R17 and R18.
    {Alt text: 2 line graphs.}}
\label{fig:appendix_velajr_2to8keV_2}
\end{figure*}

The best-fit results of the X-ray spectrum from regions R2--R18 are listed in figures \ref{fig:appendix_velajr_2to8keV_1} and \ref{fig:appendix_velajr_2to8keV_2}.

\section{Best-fit results of X-ray spectrum in the energy range of 0.5-8.0 keV}
\label{sec:appendix_05to80_fit_result}
\begin{figure*}
\begin{center}
\includegraphics[width=0.3\textwidth,clip]{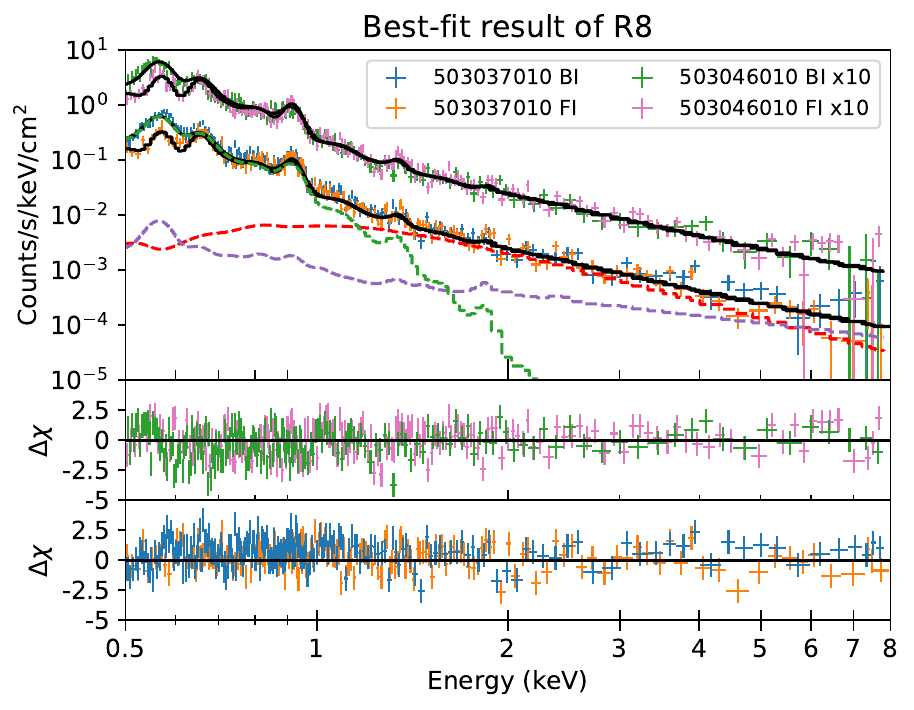}
\includegraphics[width=0.3\textwidth,clip]{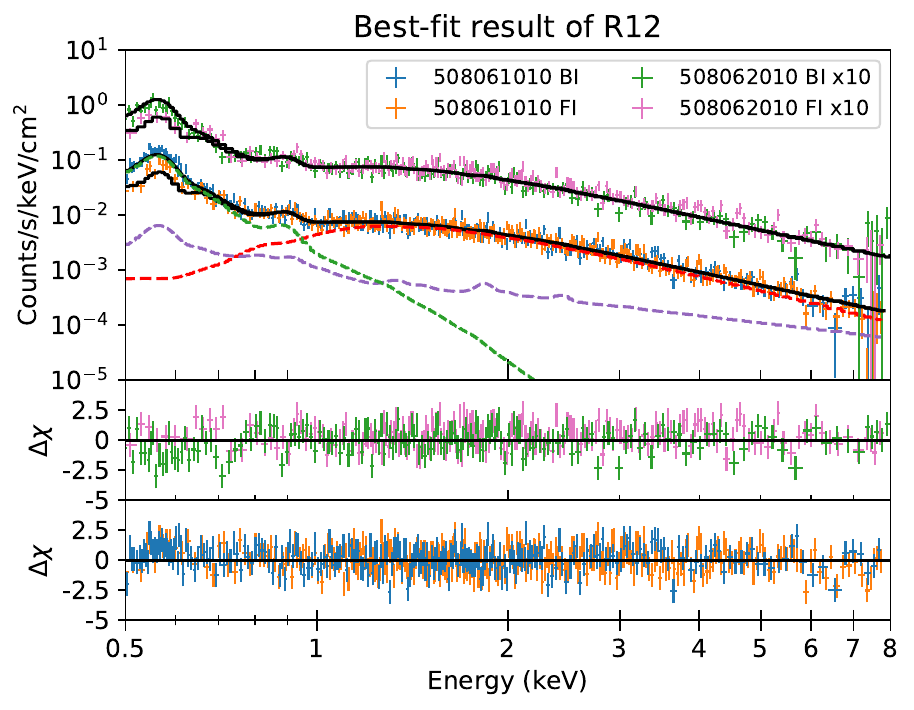}
\includegraphics[width=0.3\textwidth,clip]{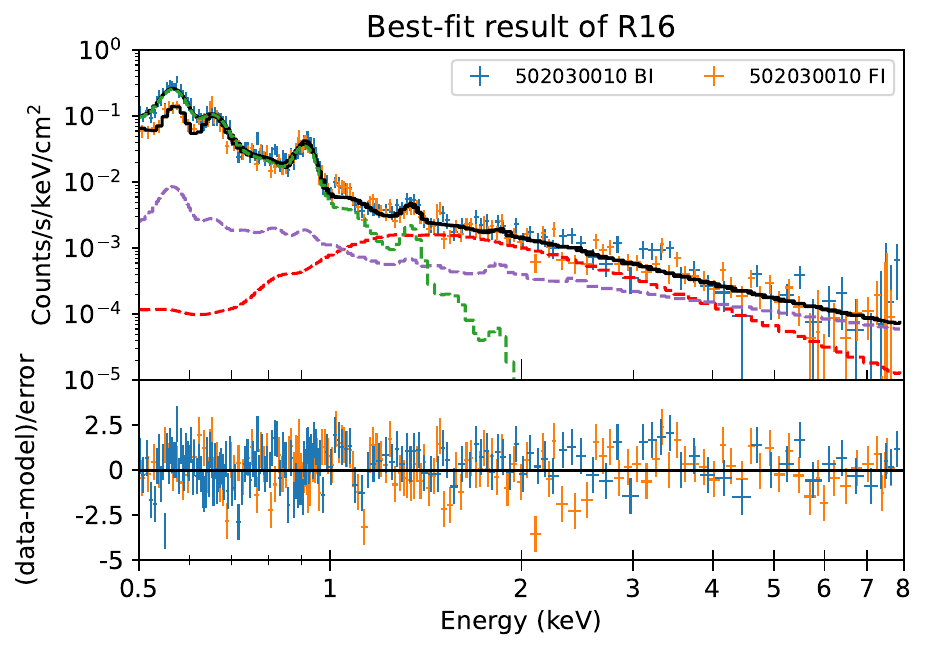}
\end{center}
\caption{
        The best-fit result of the NXB-subtracted X-ray spectrum (0.5-8.0 keV) obtained from regions R8, R12, and R16.
        The purple, red, and green dashed lines represent the X-ray background, absorbed power-law of SNR RX~J0852.0$-$4622, and absorbed thermal X-ray of Vela SNR, respectively.
        {Alt text: 3 line graphs. 
        The x-axis shows the energy from 0.5 to 8 kilo-electron volts. 
        The y-axis shows the count from 0.00001 to 10 counts per second per kilo-electron volt per square centimeter in the upper part and the residuals of minus 5 to 5 in the lower part.}
}
\label{fig:appendix_velajr_05to8keV}
\end{figure*}

The best-fit results of the X-ray spectrum from regions R8, R12, and R16 in the energy range of 0.5--8.0 keV are shown in figures \ref{fig:appendix_velajr_05to8keV}.

\section{Correlation of parameters considering the intrinsic error of X-ray flux}
\label{sec:fit_with_intscatter}
We assumed that the intrinsic error of the X-ray flux $(\sigma_{\rm int})$ is equivalent to the relative error of the cloud density (25\%), which showed the largest relative error among the correlations evaluated in our study.
We evaluated the total error of X-ray flux in each region as the square root of the sum of the squares of the statistical error and the intrinsic error $(\sigma_{\rm int} = F_{\rm X}\times 0.25)$.
We assessed the correlation between X-ray flux and (1) photon index, (2) cloud density, (3) gamma-ray counts, and (4) shock velocity using this error.
In the following, we used the same procedure described in section \ref{sec:Xflux_index} to evaluate the correlation.

\subsection{Correlation between photon index and X-ray flux}
\label{sec:fit_with_intscatter_index_flux}
\begin{figure*}
    \begin{center}
        \includegraphics[width=15cm,clip]{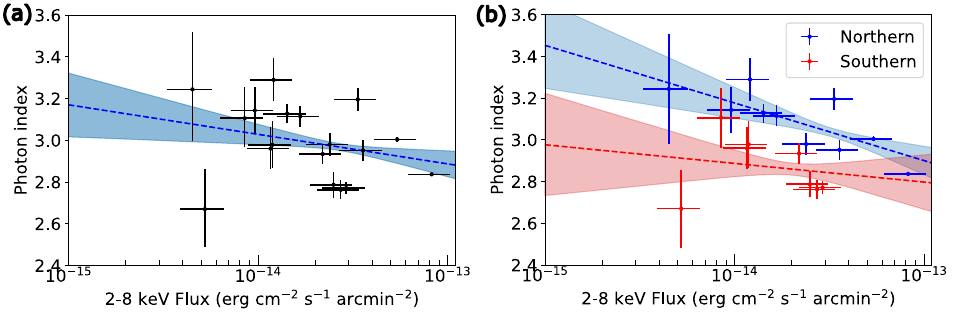}
    \end{center}
    \caption{
    Relation between the photon index and the X-ray flux considering the intrinsic error of the X-ray flux.
    Figure (a) represents the correlation plot of photon index and X-ray flux of the whole SNR.
    The best-fit results with a linear model ($\Gamma = a (\log F_{\rm X}-c) + b$) are represented in blue dashed lines, and the filled area represents 1$\sigma$ error.
    Figure (b) shows the correlation plot of the northern and southern regions, represented by blue and red dashed lines, respectively.
    We note that errors in figure (a) and (b) are $1\sigma$ confidence level.
    {Alt text: Two graphs.
    In figures (a) and (b), the x-axis shows the 2 to 8 kilo-electron volt X-ray flux from 10 to the power of minus 15 to 10 to the power of minus 13. The y-axis shows the photon index from 2.4 to 3.6.
    }
    }
    \label{fig:flux_vs_index_withinterr}
\end{figure*}

We fitted data with a logarithmic model ($\Gamma = a (\log F_{\rm X}-c) + b$) and estimated the parameters $a$ and $b$.
The resultant significance of the anti-correlation is $0.03\sigma$, $2.00\sigma$, and $0.47\sigma$ for total, northern, and southern regions, respectively.
The best-fit parameters ($a, b, c$) are presented in table \ref{tab:velajr_corelation_with_interr}.
The best-fit results are shown in blue dashed lines of figure \ref{fig:flux_vs_index_withinterr} (a) for the total region, and blue and red dashed lines for northern and southern regions, respectively of figure \ref{fig:flux_vs_index_withinterr} (b).

\subsection{Correlation between X-ray flux and cloud density}
\begin{figure*}
\begin{center}
\includegraphics[width=15cm,clip]{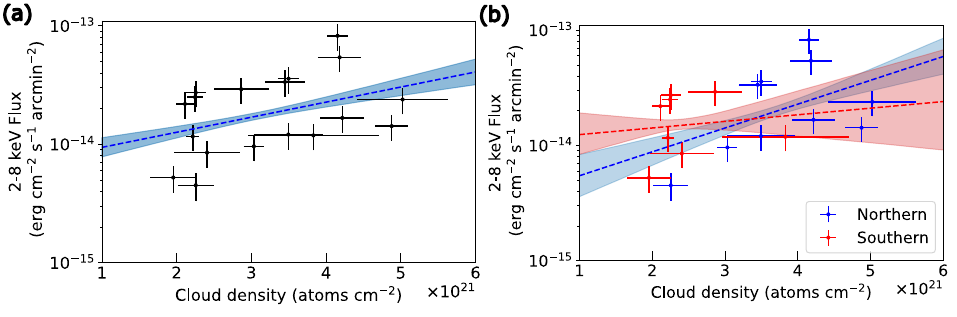}
\end{center}
\caption{
        Relation between the X-ray flux and the cloud density considering the intrinsic error of the X-ray flux.
        Figure (a) represents the correlation plot with the best-fit results of a logarithm model ($\log F_{\rm X} = a (N_{\rm cloud}-c) + b$) (blue dashed line) and 1$\sigma$ error (blue filled area).
        Figure (b) shows the correlation plot of the northern and southern regions, represented by blue and red lines, respectively.
        We note that errors in figure (a) and (b) are $1\sigma$ confidence level.
        {Alt text: Two graphs.
        In figure (a) and (b) panels, the x-axis shows the cloud density from 1 times 10 to the power of 21 to 6 times 10 to the power of 21 atoms per square centimeter.
        The y-axis shows the 2 to 8 kilo-electron volt X-ray flux from 10 to the power of minus 15 to 10 to the power of minus 13.}
}
\label{fig:flux_vs_nh_withinterr}
\end{figure*}

In order to estimate the correlation between the X-ray flux and cloud density, we fitted the data with a logarithmic model ($\log F_{\rm X} = a (N_{\rm cloud}-c) + b$) and plotted the best-fit result.
The significance of the correlation is $3.54\sigma$, $3.10\sigma$, and $0.35\sigma$ for total, northern, and southern regions, respectively.
The best-fit parameters ($a, b, c$) are presented in table \ref{tab:velajr_corelation_with_interr}.
The best-fit results are shown in blue dashed lines of figure \ref{fig:flux_vs_nh_withinterr} (a) for the total region, and blue and red dashed lines for northern and southern regions, respectively of figure \ref{fig:flux_vs_nh_withinterr} (b).

\subsection{Correlation between X-ray flux and gamma-ray counts}
\begin{figure*}
\begin{center}
\includegraphics[width=15cm,clip]{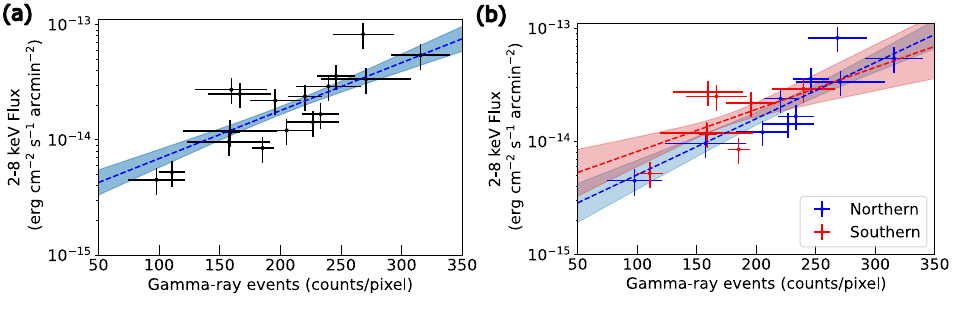}
\end{center}
\caption{
        Relation between X-ray flux and VHE gamma-ray counts (100 GeV to 100 TeV) considering the intrinsic error of the X-ray flux.
        Figures (a) and (b) represent the correlation plot of average excess VHE gamma-ray counts (100 GeV to 100 TeV) and X-ray flux.
        The blue dashed lines and blue filled area in figure (a) represent the best-fit results of a linear model ($\log F_{\rm X} = a (F_{\rm \gamma}-c) + b$) and 1$\sigma$ error, respectively.
        Figure (b) shows the correlation plot of the northern and southern regions, represented by blue and red lines, respectively.
        We note that errors in figure (a) and (b) are $1\sigma$ confidence level.
        {Alt text: Two graphs.
        In figures (a) and (b), the x-axis shows the average excess gamma-ray counts from 50 to 350 counts per pixel.
        The y-axis shows the 2 to 8 kilo-electron volt X-ray flux from 10 to the power of minus 15 to 10 to the power of minus 13.
        }
}
\label{fig:X-ray_vs_gamma-ray_withinterr}
\end{figure*}

In order to estimate the correlation between the X-ray flux and gamma-ray flux, we fitted the data with a logarithmic model ($F_{\rm \gamma} = a (\log F_{\rm X}-c) + b$) and plotted the best-fit result.
The significance of the correlation is $5.66\sigma$, $5.20\sigma$, and $2.26\sigma$ for total, northern, and southern regions, respectively.
The best-fit parameters ($a, b, c$) are presented in table \ref{tab:velajr_corelation_with_interr}.
The best-fit results are shown in blue dashed lines of figure \ref{fig:X-ray_vs_gamma-ray_withinterr} (a) for the total region, and blue and red dashed lines for northern and southern regions, respectively of figure \ref{fig:X-ray_vs_gamma-ray_withinterr} (b).

\subsection{Correlation between the averaged shock velocity and X-ray flux}
\label{sec:fit_with_intscatter_velo_flux}
\begin{figure*}
\begin{center}
\includegraphics[width=15cm,clip]{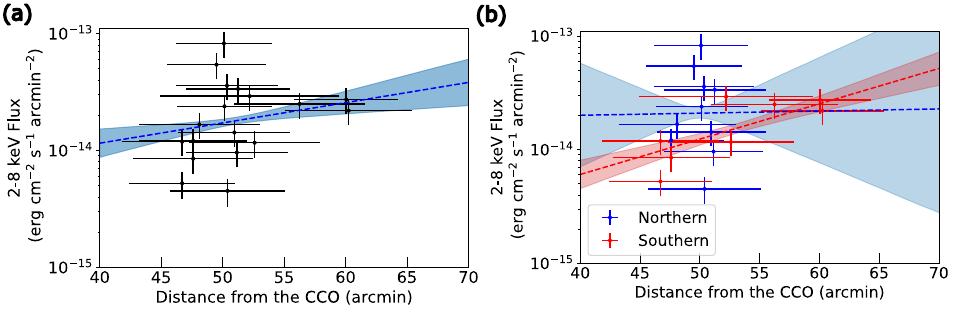}
\end{center}
\caption{
        Relation between X-ray flux and shock velocity considering the intrinsic error of the X-ray flux.
        Figure (a) represents the correlation plot with best-fit results of a linear model ($\log F_{\rm X} = a (D_{\rm CCO}-c) + b$) (blue dash line) and 1$\sigma$ error (blue filled area).
        Figure (b) shows the correlation plot of the northern and southern regions, represented by blue and red lines, respectively.
        We note that errors in figure (a) and (b) are $1\sigma$ confidence level.
        {Alt text: Two graphs.
        In figures (a) and (b), the x-axis shows the distance from the CCO from 40 to 70 arc-minutes.
        The y-axis shows the 2 to 8 kilo-electron volt X-ray flux from 10 to the power of minus 15 to 10 to the power of minus 12.}
}
\label{fig:Xflux_vs_distance_withinterr}
\end{figure*}

In order to estimate the correlation between the X-ray flux and shock velocity, we fitted the data with a logarithmic model ($\log F_{\rm X} = a (D_{\rm CCO}-c) + b$) and plotted the best-fit result.
The significance of the correlation is $1.60\sigma$, $0.04\sigma$, and $3.40\sigma$ for total, northern, and southern regions, respectively.
The best-fit parameters ($a, b, c$) are presented in table \ref{tab:velajr_corelation_with_interr}.
The best-fit results are shown in blue dashed lines of figure \ref{fig:Xflux_vs_distance_withinterr} (a) for the total region, and blue and red dashed lines for northern and southern regions, respectively of figure \ref{fig:Xflux_vs_distance_withinterr} (b).

\begin{table*}
\tbl{Correlation coefficient between parameters and best-fit parameters considering the intrinsic error of the X-ray flux\footnotemark[$*$]}{
\begin{tabular}{cccccccc}
\hline
  \multicolumn{3}{c}{Parameter} & Region & Significance of correlation & \multicolumn{3}{c}{Best-fit parameter} \\
  X-axis & & Y-axis & & $\sigma$ & $a \pm \Delta a$ & $b \pm \Delta b$ &  $c$ \\
\hline
X-ray flux       & vs. & Photon index   & Total     & 0.03       & $(-1.4 \pm 1.1) \times 10^{-1}$      & $2.99\pm0.02$      & $-13.743$                \\
                 &     &                & Northern  & 2.00       & $(-2.8 \pm 1.4) \times 10^{-1}$      & $3.09\pm0.03$      & $-13.680$                \\
                 &     &                & Southern  & 0.47       & $(-9 \pm 19) \times 10^{-2}$         & $2.87\pm0.04$      & $-13.822$                \\\hline
Cloud density    & vs. & X-ray flux     & Total     & 3.54       & $(1.3 \pm 0.4) \times 10^{-22}$      & $-13.74\pm 0.03$   & $3.22 \times 10^{21}$ \\ 
                 &     &                & Northern  & 3.10       & $(2.1 \pm 0.7) \times 10^{-22}$      & $-13.68\pm 0.04$   & $3.8 \times 10^{21}$ \\ 
                 &     &                & Southern  & 0.35       & $(0.6 \pm 1.7) \times 10^{-22}$      & $-13.82\pm 0.04$   & $2.5 \times 10^{21}$ \\\hline 
Gamma-ray counts & vs. & X-ray flux     & Total     & 5.66       & $(4.2 \pm 0.7) \times 10^{-3}$       & $-13.74\pm 0.03$   & $201$                        \\
                 &     &                & Northern  & 5.20       & $(5.0 \pm 0.1) \times 10^{-3}$       & $-13.68\pm 0.04$   & $224$                        \\
                 &     &                & Southern  & 2.26       & $(3.7 \pm 1.6) \times 10^{-3}$       & $-13.82\pm 0.04$   & $172$                        \\\hline
Shock velocity   & vs. & X-ray flux     & Total     & 1.60       & $(1.7 \pm 1.1) \times 10^{-2}$       & $-13.74\pm0.03$    & $51.2$                     \\
                 &     &                & Northern  & 0.04       & $(2 \pm 49) \times 10^{-3}$          & $-13.68\pm0.04$    & $49.9$                     \\
                 &     &                & Southern  & 3.40       & $(3.1 \pm 0.9) \times 10^{-2}$       & $-13.82\pm0.02$    & $53$                     \\
\hline
\end{tabular}}\label{tab:velajr_corelation_with_interr}  
\begin{tabnote}
\footnotemark[*] The errors represent a $1\sigma$ confidence level on an interesting single parameter.  \\ 
\end{tabnote}
\end{table*}

\subsection{Effects of intrinsic error of X-ray flux}
The results of sections A.\ref{sec:fit_with_intscatter_index_flux} to A.\ref{sec:fit_with_intscatter_velo_flux} indicated a lower significance of the correlation between parameters when the intrinsic error was taken into account.
However, parameters that had strong correlations with significance greater than 3$\sigma$ before accounting for intrinsic error remained significant at more than 3$\sigma$ after the consideration.
Furthermore, parameters with correlations with significance of less than 3$\sigma$ in the analysis of section \ref{sec:azimuth_dist_parameters} did not reach the 3$\sigma$ confidence level after considering intrinsic error.
From these results, we conclude that the intrinsic error of the X-ray flux does not affect our conclusions.

\clearpage

\bibliographystyle{apj.bst}
\bibliography{sample631.bib}

\end{document}